\documentclass[prx,twocolumn,floatfix,footinbib,bibnotes,longbibliography]{revtex4-1}

\usepackage{graphicx}
\usepackage{epsfig,tabularx,multirow,booktabs}
\usepackage{amsmath,bbm,amssymb,bm,times}
\usepackage[colorlinks,linkcolor=blue,citecolor=blue]{hyperref}

\newcommand{\eqn}[1]{
\begin{eqnarray}
	#1
\end{eqnarray}
}

\newcolumntype{C}{>{$}c<{$}}
\AtBeginDocument{
\heavyrulewidth=.08em
\lightrulewidth=.05em
\cmidrulewidth=.03em
\belowrulesep=.65ex
\belowbottomsep=0pt
\aboverulesep=.4ex
\abovetopsep=0pt
\cmidrulesep=\doublerulesep
\cmidrulekern=.5em
\defaultaddspace=.5em
}

\begin{document}
\title{Quantum Electrodynamic Control of Matter: Cavity-Enhanced Ferroelectric Phase Transition}
\author{Yuto Ashida}
\email{ashida@ap.t.u-tokyo.ac.jp}
\affiliation{Department of Applied Physics, University of Tokyo, 7-3-1 Hongo, Bunkyo-ku, Tokyo 113-8656, Japan}
\author{Ata\c c $\dot{\mathrm{I}}$mamo$\breve{\mathrm{g}}$lu}
\affiliation{Institute of Quantum Electronics, ETH Zurich, CH-8093 Z{\"u}rich, Switzerland}
\author{J\'er\^ome Faist}
\affiliation{Institute of Quantum Electronics, ETH Zurich, CH-8093 Z{\"u}rich, Switzerland}
\author{Dieter Jaksch}
\affiliation{Clarendon Laboratory, University of Oxford, Parks Road, Oxford OX1 3PU, United Kingdom}
\author{Andrea Cavalleri}
\affiliation{Max Planck Institute for the Structure and Dynamics of Matter, 22761 Hamburg, Germany}
\affiliation{Clarendon Laboratory, University of Oxford, Parks Road, Oxford OX1 3PU, United Kingdom}
\author{Eugene Demler}
\affiliation{Department of Physics, Harvard University, Cambridge, MA 02138, USA}

\begin{abstract} 
The light-matter interaction can be utilized to qualitatively alter physical properties of materials. Recent theoretical and experimental studies have explored this possibility of controlling matter by light based on driving many-body systems via strong classical electromagnetic radiation, leading to a time-dependent Hamiltonian for electronic or lattice degrees of freedom.  
To avoid inevitable heating, pump-probe setups with ultrashort laser pulses have so far been used to study transient light-induced modifications in materials.
Here, we pursue yet another direction of controlling quantum matter by modifying quantum fluctuations of its electromagnetic environment. 
In contrast to earlier proposals on light-enhanced electron-electron interactions, we consider a dipolar quantum many-body system embedded in a cavity composed of metal mirrors, and formulate a theoretical framework to manipulate its equilibrium properties on the basis of quantum light-matter interaction. 
We analyze hybridization of different types of the fundamental excitations, including dipolar phonons, cavity photons, and plasmons in metal mirrors, arising from the cavity confinement in the regime of strong light-matter interaction. This hybridization qualitatively alters the nature of the collective excitations and can be used to selectively control energy-level structures in a wide range of platforms. Most notably, in quantum paraelectrics, we show that  the cavity-induced softening of infrared optical phonons enhances the  ferroelectric phase in comparison with the bulk materials. Our findings suggest an intriguing possibility of inducing a superradiant-type transition via the light-matter coupling without external pumping. We also discuss possible applications of the cavity-induced modifications in collective excitations to molecular materials and excitonic devices.
\end{abstract}

\maketitle

\section{Introduction}
One of the central goals in both quantum optics and condensed matter physics is to 
understand macroscopic phenomena emerging from the fundamental quantum degrees of freedom.
 Historically, quantum optics strived to study light-matter interactions mainly in few-body regimes \cite{CCT89}. On another front, condensed matter physics investigated collective phenomena of many-body systems, and  largely focused on optimizing structural and electronic phases of solids. In particular, a number of techniques, including applied strain, electronic doping, and isotope substitution, have been used to enhance a macroscopic order such as ferroelectricity or superconductivity. As an interdisciplinary frontier at the intersection of optics and condensed matter physics, there have recently been remarkable developments in using classical electromagnetic radiation to control transient states of matter \cite{KT13}. The aim of this paper is to explore yet another route towards controlling the phase of matter by {\it quantum} light, namely, by modifying quantum fluctuations of the electromagnetic field in equilibrium -- in the absence of an external drive.

Studies of light-induced modifications in material properties date back to experiments by  Dayem and Wyatt \cite{DAH64,WAFG66}, who observed that coherent microwave radiation leads to an increase of the critical current in superconductors.  This phenomenon has been understood from the Eliashberg electron-photon theory \cite{EGM70,IBI73,AS77}, in which photons with frequencies below the quasiparticle threshold are shown to modify the distribution of quasiparticles in a way that the superconducting order parameter is enhanced.  
 Other seminal works studying control of matter by light include Floquet engineering of electronic band structures \cite{OT09,KT11,LN11,WYH13,MJW20}, and photo-induced pumping by ultrashort laser pulses for creating metastable nonequilibrium states   \cite{RM07,FM11,MR14,MR142,KAF15,MM16,PE17,PAA16,KM16,KA16,SMA17,SMA172,KDM17,TDN18,ZJ14,AC18}. 
Nonlinear optical phenomena can also be understood from the perspective of light-induced changes in optical properties of matter \cite{PH19,CA18,SY16}.
For instance, the light-induced transparency can be considered as strong modification of the refractive index and absorption coefficient by a control optical beam \cite{FM05}, and the optical phase conjugation phenomena arise from the resonant parametric scattering of signal photons and matter excitations \cite{ZB85,SS19}. 
The essential element of all these developments is an external pump that provides classical light field acting on quantum matter.  

An alternative approach to controlling matter by light is to utilize quantum fluctuations of the vacuum electromagnetic field. A prototypical model to understand such quantum light-matter interaction is the Dicke model, which describes an ensemble of two-level atoms coupled to a single electromagnetic mode of a cavity \cite{DRH54,TM68,KH73,HFT73,EC032,EC03}. For a sufficiently strong electric dipole, Dicke has shown that one should find instability into the superradiant phase characterized by spontaneous atomic polarization and macroscopic photon occupation \cite{DRH54}. This transition originates from the competition
between the decrease of the total energy due to the light-matter coupling and the energy cost of adding photons to the cavity and admixing the excited state of atoms. The Dicke model is exactly solvable via the Bethe-ansatz equations originally discussed in the central spin model  \cite{MG76,RWR63,TO10}, and the existence of the superradiant transition has been rigorously proven in the thermodynamic limit \cite{KH73,HFT73}.

Subsequently to Dicke's original work, it has been pointed out that the superradiant transition requires such strong interaction that one needs to include the $\hat{A}^2$ term in the Hamiltonian, where $\hat{A}$ denotes the vector potential of electromagnetic fields. In turn, the latter term has been shown to prevent the transition \cite{RK75,NP10}.
Even though ideas have been suggested to circumvent this no-go theorem by using microwave resonators \cite{CG07,KJ09} or including additional inter-particle interactions  \cite{JK07,GT16,JT16,DB18,MG19,LK20}, the problem is still a subject of debate \cite{CL12,DB182,AGM19,AGM20,SA19}. 
An alternative proposal using multilevel systems with external pump fields  \cite{DF07,TEG13} has been experimentally realized \cite{Zhiqiang:17,ZZ18}. Analogues of the Dicke transition have also been explored in ultracold atoms, where  matter excitations correspond to different momentum states of atoms  \cite{MC08,BK10} or hyperfine states in spin-orbit-coupled Bose gases \cite{HC14}. In all these realizations, external driving is essential to obtain the superradiant-type phases.

\begin{figure}[t]
\includegraphics[width=70mm]{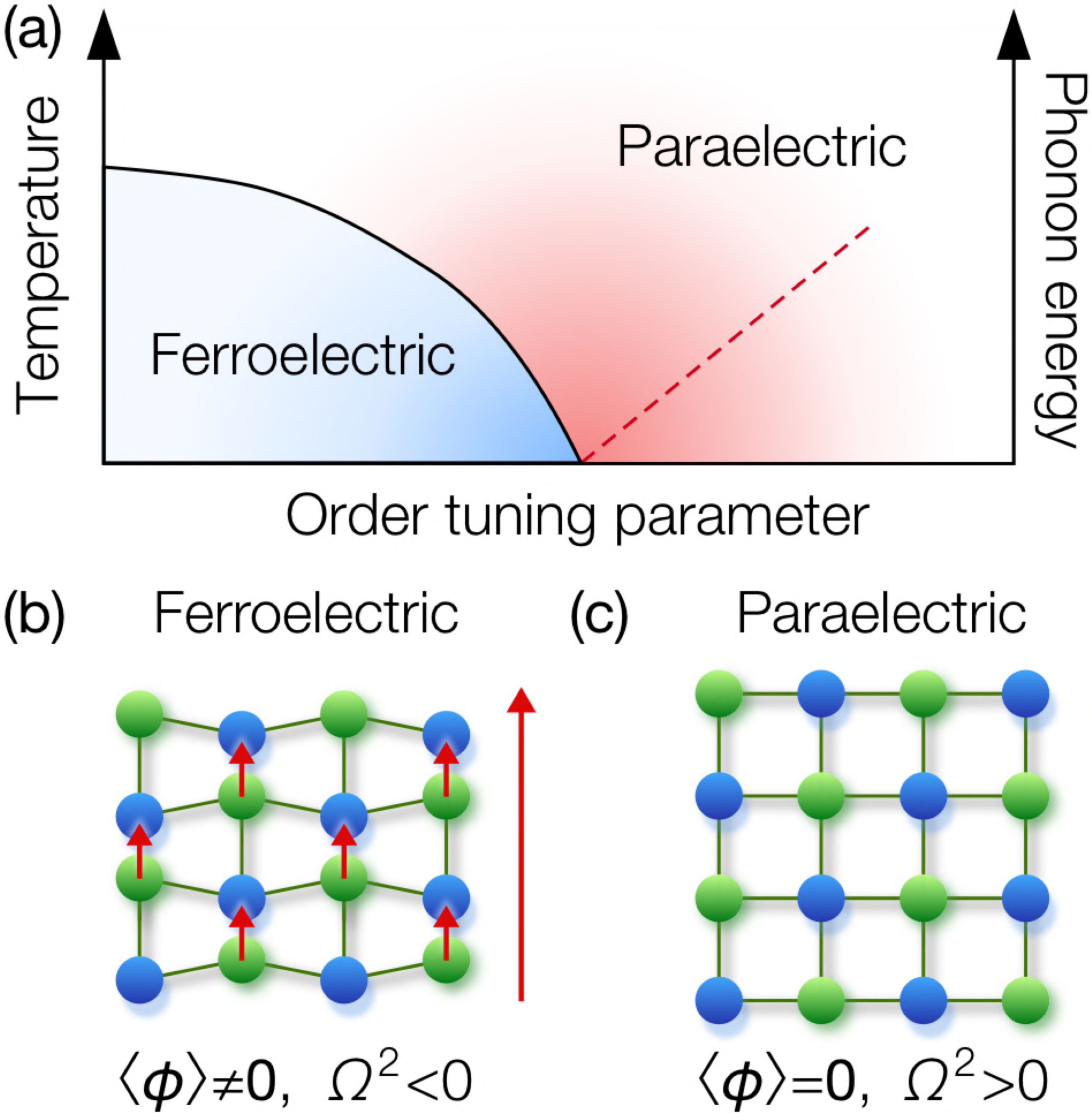} 
\caption{\label{fig1}
(a) Schematic phase diagram of a quantum paraelectric material. The horizontal axis corresponds to a tuning parameter, such as pressure or isotope substitution, which controls the transition between the ferroelectric and paraelectric phases. The vertical axis is temperature/energy and the red dashed line in the paraelectric phase shows the optical phonon energy. Note that the phonon softens to zero energy at the paraelectric-to-ferroelectric transition point.
The main goal of this paper is to demonstrate that this transition can also be  controlled by changing the electromagnetic environment of the system, in particular, by placing a thin film of material in a cavity.
(b,c) Ferroelectric and paraelectric phases in the case of ionic crystals. Green (blue) circles indicate the ions with the positive (negative) charges. 
The ferroelectric phase in (b) is characterized by a nonzero uniform displacement causing  an electric polarization (red arrows) while the mean-phonon displacement is absent in the paraelectric phase in (c). The variables $\boldsymbol\phi$ and $\Omega$ represent the phonon field and the bare phonon frequency, respectively (see Sec.~\ref{sec3}). 
}
\end{figure}

\begin{figure}[t]
\includegraphics[width=70mm]{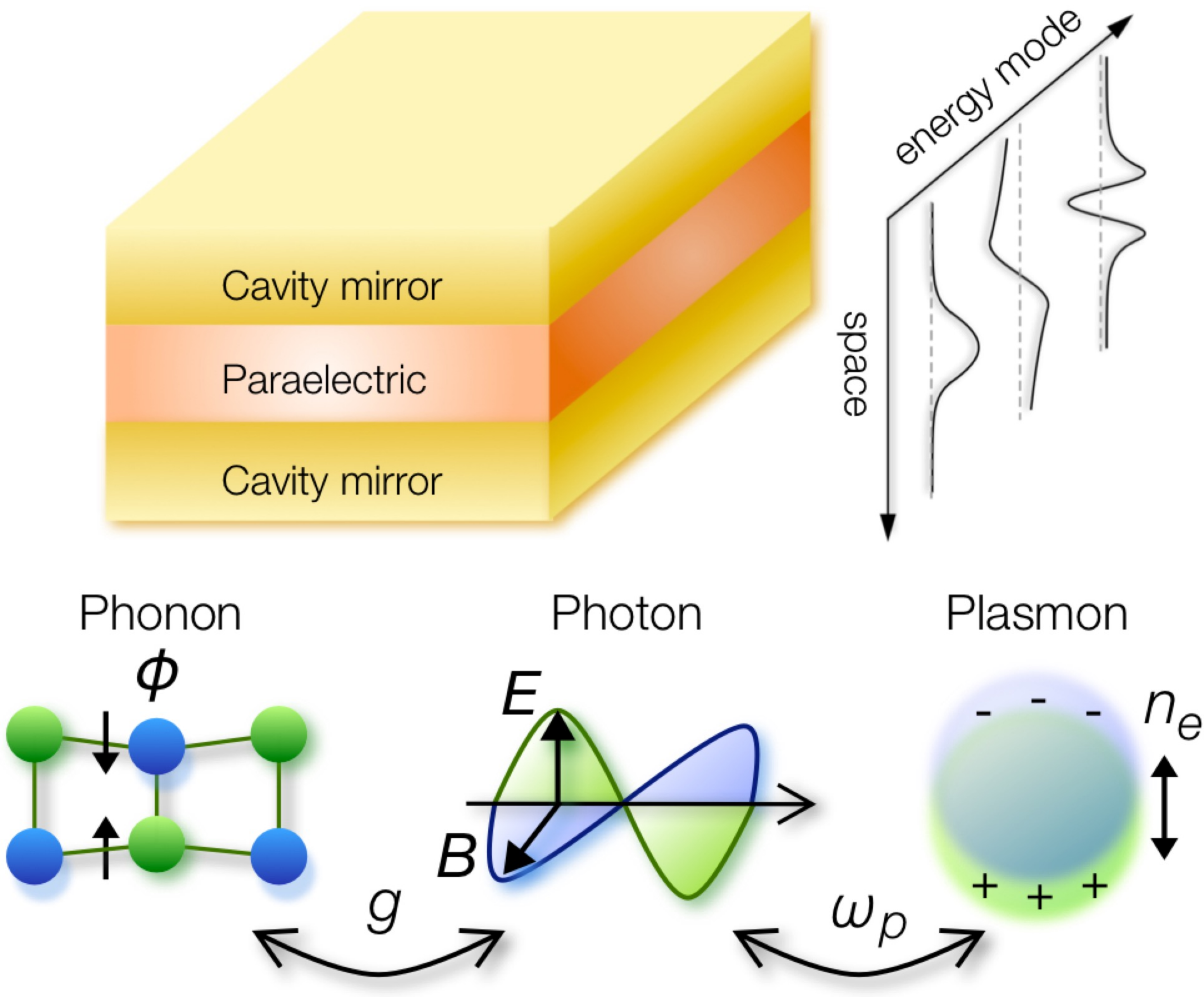} 
\caption{\label{fig2}
(Top) Setup of the heterostructure cavity configuration considered in the present manuscript. A slab of paraelectric material with a finite thickness is sandwiched between two semi-infinite metallic electrodes. Right panel represents spatial profiles of typical energy modes along the spatial direction perpendicular to the metal-insulator interfaces.
(Bottom) Schematic figure illustrating three fundamental excitations relevant to the theory. A phonon excitation describes an optical collective mode induced by, e.g., the deformation of ionic crystals around the equilibrium configuration. More generally, it can also represent dipole moments of molecular solids or assembled organic molecules. The interaction between phonons and cavity photons is mediated by the dipole coupling, whose strength is represented by $g$. A plasmon, a collective excitation of electronic gases in metal mirrors, leads to the electromagnetic coupling characterized by the plasma frequency $\omega_p$.
}
\end{figure}

The present work in this context suggests a promising route towards realizing a genuine equilibrium superradiant transition under no external drive, which has so far remained elusive. Our consideration is motivated by an observation that transition into a superradiant state should be easier to attain in a system naturally close to a phase with spontaneous electric polarization. For instance, materials such as SrTiO$_3$ and KTaO$_3$ remain paraelectric down to zero temperature, but ferroelectric phase can be induced by applying pressure or isotope substitution. The primary goal of this paper is to investigate cavity-induced changes in a quantum paraelectric at the verge of the ferroelectric order (see Fig.~\ref{fig1}(a-c)). 
We formulate a simple yet general theoretical framework for analyzing hybridization of matter excitations, such as infrared active phonons, and light modes confined in systems where metallic cladding layers act as cavity mirrors (see Fig.~\ref{fig2}). 
Resulting polariton excitations consist of infrared active phonons in the quantum paraelectric, electromagnetic fields in the cavity, and plasmons in the metallic electrodes. We show that the spectrum of such cavity polaritons is qualitatively different from the bulk spectrum, and can exhibit significant mode-softening, indicating  enhancement of the ferroelectric instability.

Two essential elements of our analysis place it outside the recently discussed no-go theorems for the superradiant transition \cite{RK75,NP10,DB182,AGM19,SA19}. Firstly, we include multiple modes for both light and matter degrees of freedom. More precisely, we assume systems to be large enough in the direction parallel to the interfaces in such a way that a continuum of spatially varying electromagnetic fields and phonon excitations must be included. This should be contrasted to most of the previous studies of the superradiant transitions, which have focused on a simplified case of either many-body systems coupled to a single spatially uniform mode in a cavity, or  two-level systems coupled to a single mode obtained after truncating high-lying spatially varying modes \cite{JK07}. The second key element of our analysis is including nonlinearity of dipolar phonons in materials. In the bulk geometry, the importance of such nonlinearities for understanding the quantum phase transition has been discussed in Ref.~\cite{PL09}. In contrast, we focus on the cavity geometry and  demonstrate that a combination of nonlinearities and cavity confinement leads to substantial phonon softening. This softening in turn enhances the superradiant transition, thus opening a way to controlling quantum matter by modifying electromagnetic vacuum.

Before concluding this section, we put our work in a broader context of studies on quantum matter strongly interacting with electromagnetic fields. Importantly, our model bears a close connection with physical systems discussed in several areas, including cavity quantum electrodynamics (QED), plasmonics, and polaritonic chemistry.
Firstly, in the field of cavity QED, one is typically interested in small quantum systems coupled to cavity photons \cite{PEM46,DRH54,ETJ63,HJK98,RJM01,RM05,HW06}; prominent examples include individual atoms \cite{BA04,BKM05,HC17}, quantum dots  \cite{KG06,RJP04,YT04,HK07,RMT09,LG15}, and light-emitting defects in solids such as nitrogen- and silicon-vacancy centers \cite{CHS08,FA12,AR13,LL15,RMJ15,SA16,RD17}. 
These systems have been suggested as platforms for realizing single-photon transistors \cite{LP15,HB16,SS18} and sources \cite{AI16,DX16,SH18,DS19}, and can thus provide a promising route to enabling photonic quantum information processing \cite{ZSB00,DLM04,AM14}. 
Achieving a strong light-matter coupling has also been the subject of intense research in the fields of plasmonics in nanostructures \cite{BW03,JMP06,TMS13} and polaritonic chemistry \cite{HJA12,GJ15,ETW16,FJ15,FJ172,FJ17}. In both areas, cavity setups hold promise for realizing systems with a broad range of interesting physical properties, including superconductivity \cite{SMA18,SF19,CJB19}, charge and energy transport \cite{OE15,FeJ15,SJ15,HD17,ZX17,HD18,PBGL19,LJ20}, hybridized excitations in molecular crystals \cite{HRJ04,KCS10,CT16,SK18} and  light-harvesting complexes \cite{CDM14,MMLA19,Eiznereaax4482,C9SC04950A}, chemical reactivity of organic compounds \cite{TA16,TA19,MMLA18}, and energy transfer via phonon nonlinearity \cite{JDM19}. Strong coupling between the zero-point fluctuations of the electromagnetic field (i.e., the vacuum fluctuations) and vibrational excitations of individual molecules has been experimentally realized \cite{CR16}, and its nontrivial influence on the superconductivity has recently been reported \cite{TA192}.

In this context, one of the main novelties of the present work is a general formalism that allows one to include both the dispersive nature of polariton excitations in the dielectric slab of finite thickness and the surface plasmonic effects occurring at the metal-insulator interfaces, i.e., a finite plasma frequency of the metallic cavity mirrors.
 We demonstrate that the proper consideration of these aspects is important to accurately understand several key features in the strong light-matter coupling regime, which are not readily attainable in the conventional bulk settings. 
The resulting cavity-induced changes  provide a general route towards controlling energy-level structures not only in paraelectric materials, but also in many other systems such as molecular solids, organic light-emitting devices, and excitonic systems.
Our analysis can further be extended to include fabricated surface nanostructures \cite{YH19,BJJ19}.
In view of the versatility of the present formulation, we anticipate that our results will be useful for advancing our understanding of a variety of physical systems lying at the intersection of condensed matter physics, quantum optics, and quantum chemistry.

The remainder of the paper is organized as follows. In Sec.~\ref{sec2}, we  summarize our main results at a nontechnical level. In Sec.~\ref{sec3}, we present a theoretical framework for describing hybridized collective modes in both bulk and cavity settings. As a first step, we neglect phonon nonlinearities and determine  dispersion relations of elementary excitations.  In Sec.~\ref{sec4}, we include phonon nonlinearities based on the  variational approach and analyze their effects on the paraelectric-to-ferroelectric phase transition. We show that the mode confinement in the cavity geometry induces significant softening of polariton modes, indicating the enhanced  propensity for ferroelectricity. 
In Sec.~\ref{sec5}, we present a summary of results and suggest several interesting directions for future investigations. 

\section{Summary of main results\label{sec2}}
  \begin{figure}[b]
\includegraphics[width=70mm]{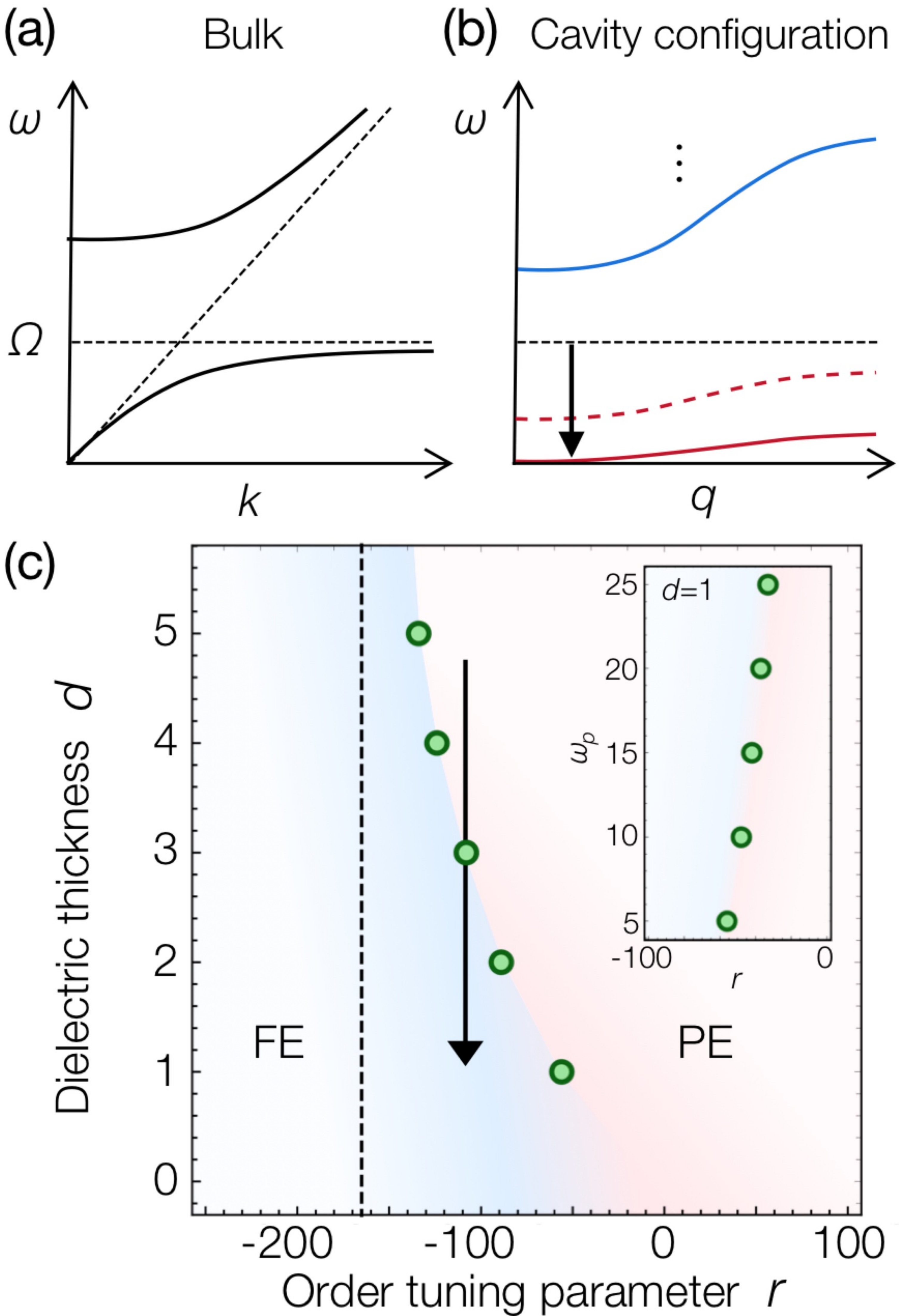} 
\caption{\label{fig3}
 Summary of the cavity-induced changes of the collective excitations and the phase diagram. (a) Bulk band dispersions. The hybridization between a photon  and an optical phonon at the frequency $\Omega$ (black dashed lines) leads to two gapped branches corresponding to the upper and lower transverse phonon polaritons (black solid curves).  (b) Collective excitations confined in the cavity as a function of the in-plane wavevector $q$. The hybridization leads to softening of the optical phonon modes (red curves and black solid arrow) and also supports the coupled surface modes (blue curve). (c) Phase diagram in the cavity geometry. As the tuning parameter $r$ is decreased, the transition from the paraelectric phase (PE) to the ferroelectric phase (FE) occurs when the renormalized effective phonon frequency goes down to zero.   
The FE is enhanced in the cavity setting in comparison with the bulk geometry, for which the transition point is indicated by the black dashed line. The inset shows the dependence on the plasma frequency $\omega_p$ at the insulator thickness $d=1$.  The parameters are $c=\hbar=\beta=1$, $\omega_p=5$, $U=1$,  $g=3.5$, and the UV momentum cutoff is $\Lambda_c=10^2$.
}
\end{figure}
In this section, we summarize the main results of the paper at a nontechnical level before presenting a detailed theoretical formulation in subsequent sections. 
Our most important contribution is a theoretical proposal of a new approach for controlling many-body states of matter using quantum light that does not rely on external pumping. 
 Physical phenomenon that underlies this proposal is the cavity-induced softening of  a dipolar phonon mode, which originates from the interplay between phonon nonlinearities and modifications of the hybrid light-matter collective excitations in the cavity geometry (see Fig.~\ref{fig3}(a,b)). Softening  of the phonon frequency to zero  indicates a phase transition into the state with spontaneous symmetry breaking and formation of electric dipolar moments (red dashed line in Fig.~\ref{fig1}(a)). Thus, our analysis demonstrates that  cavity confinement shifts the transition point in favor of the ferroelectric phase, i.e., the paraelectric-to-ferroelectric transition  occurs at a lower applied pressure/isotope substitution in the cavity geometry as compared to the bulk case (Fig.~\ref{fig3}(c)).  Said differently,  it is possible to start with a material that is in the symmetry-unbroken paraelectric phase in a bulk sample, yet will be in the ferroelectric phase when placed inside the cavity (black downarrow in Fig.~\ref{fig3}(c)).

The main focus of this paper is on many-body systems that are naturally close to the superradiant-type transition, i.e., paraelectric materials close to the ferroelectric phase. 
For instance, SrTiO$_3$ and KTaO$_3$ are paraelectric at ambient conditions, but isotope substitution of $^{18}$O for $^{16}$O or applying pressure can bring the materials into a phase with spontaneous electric polarization \cite{RSE14}. The key property of such materials relevant to our analysis is the existence of a low-energy infrared optical phonon mode. To understand collective modes of such  systems in resonator settings shown in Fig.~\ref{fig2}, we need to consider hybridization of the electric dipole moment of the optical phonon with the quantized electromagnetic modes confined in the cavity and with plasmons in metal mirrors.

Our analysis of quantum light-matter coupling has implications for understanding both quantum fluctuations at zero temperature and thermal fluctuations at finite temperature. Placing the focus on equilibrium systems makes the present work distinct from the previous studies on materials subject to classical light waves, in which the use of strong external drive is essential. 
It is such {\it quantum} aspect of light-matter interaction that motivates us to ask the two questions:
\begin{itemize}
\item[(A)]{How does the light-matter coupling modify polariton modes of paraelectric materials in the cavity geometry?}
\item[(B)]{Is it possible to enhance the ferroelectric instability by modifying  quantum electromagnetic environment of a many-body system under no external pump?}
\end{itemize}

To address these questions, we formulate a simple yet generic theoretical framework for analyzing interaction between three types of excitations: infrared active phonons, cavity photons, and plasmons in metal mirrors.  We focus on the heterostructure, in which energies can be confined in volumes much smaller than the free-space wavelengths of photons, leading to strong enhancement of light-matter interaction \cite{CDE07}. 
Strong hybridization between the continuum of electromagnetic fields and matter excitations necessitates deriving new collective modes starting from the microscopic model and not relying on modes defined in the bulk geometry. 

In our analysis, the size of the cavity in the longitudinal direction is assumed to be large enough for collective  eigenmodes to be characterized by a conserved in-plane momentum. Thus, the present model is distinct from most of the previous studies on the superradiant phases, where only one or at most a few  modes in a cavity have been included. In the ultrastrong coupling regime, it is essential to accurately take into account the full {\it continuum} of modes when analyzing thermodynamic properties of the system. The careful inclusion of all the proper modes becomes particularly important when the phonon nonlinearity is included because, as discussed below, the frequency of every mode is affected by quantum and thermal fluctuations of the other modes (see Sec.~\ref{sec4} for further details).

In view of the importance of  identifying the correct cavity eigenmodes, 
 we begin our analysis with a detailed discussion of the quadratic theory to answer the first question (A). Interestingly, even when the system is far from the transition point,  the cavity confinement can qualitatively alter the nature of collective excitations (see e.g., Fig.~\ref{fig6} in Sec.~\ref{sec3}). Building upon this, we proceed to address the second question (B) and show that, after including  phonon nonlinearities in the analysis, the cavity polaritons can exhibit considerable mode softening in comparison to their analogues in bulk materials. This softening ultimately induces the paraelectric-to-ferroelectric phase transition when the frequency of the lowest energy mode goes to zero (see Fig.~\ref{fig3}(b)).

We note that the phenomenon of cavity enhancement of the ferroelectric phase is not present at the level of a quadratic model, and can only be found when phonon nonlinearities are included in the analysis. The phonon nonlinearities in general render the formation of a spontaneous dipole moment energetically unfavorable. 
We show that the cavity confinement alleviates this adverse effect of phonon-phonon interactions, thus allowing one to induce ferroelectricity in quantum paraelectrics  that otherwise exhibit no spontaneous polarizations even at zero temperature.  We demonstrate that the phenomenon of cavity-enhanced ferroelectricity is more pronounced when confinement effects are enhanced by using a cavity with a thinner slab of paraelectric material and/or metallic mirrors with a higher plasma frequency (see Fig.~\ref{fig3}(c)).

A few remarks are in order.  While our theory is formulated for a heterostructure based on a paraelectric sandwiched between metallic mirrors, the present approach of controlling material excitations via modifying quantum electromagnetic environment can readily be generalized to a wide range of platforms. To demonstrate this point, we discuss several other potential applications of such cavity-based control of polariton modes, including molecular materials and excitonic devices. Thus, our analysis may find practical applications in improving the efficiency of organic light-emitting devices and realizing photon-exciton converters (see Sec.~\ref{sec5} and App.~\ref{app4}). We also consider the role of dissipation on the polariton softening and phase transition using a standard Drude-type model of conductivity in metallic mirrors. Our formalism can be extended to include other mechanisms of dissipation.

\section{Phonon-photon-plasmon Hybridization\label{sec3}}
Our main goal is to understand the qualitative physics of a dipolar insulator coupled to quantized electromagnetic fields in a cavity rather than to make predictions specific to particular experimental setups. Thus, we consider a generic microscopic model and will provide order of magnitude estimates for a concrete system at the end of our analysis in Sec.~\ref{sec4}. We here begin by introducing a general theory of the light-matter interaction and formulate a framework for obtaining hybridized eigenmodes in a way that allows us to quantize the fields. Throughout this paper, we will assume that phonon frequencies are the same for all phonon polarizations, while generalization of our analysis to anisotropic systems may introduce additional interesting features (see e.g., Ref.~\cite{GA18}). In the next section, we will apply the framework to an ionic crystal confined in the cavity as a concrete case and analyze its paraelectric-to-ferroelectric phase transition.  

\subsection{Bulk}
Before moving to the case of the cavity setting, we first discuss the three-dimensional bulk setup with no boundary effects and determine the bulk dispersion relations for the phonon-photon hybridized modes, i.e., the phonon polaritons. Later, we generalize the theory to the cavity configuration by taking into account  confinement effects and also contributions from plasmons in metal mirrors.

The Hamiltonian of the bulk system is given by
\eqn{
\hat{H}_{{\rm tot}}=\hat{H}_{{\rm light}}+\hat{H}_{{\rm matter}}+\hat{H}_{{\rm l{\rm -}m}},
}
where the first (second) term describes the dynamics of the electromagnetic field (dipolar matter field), and the last term represents the light-matter coupling between these two fields. We specify each of those terms below.

\subsubsection{Electromagnetic field}
The free evolution of the electromagnetic field is described by the Hamiltonian
\eqn{\label{emfree}
\hat{H}_{{\rm light}}=\int d^{3}r\left[\frac{\hat{\boldsymbol{\Pi}}^{2}}{2\epsilon_{0}}+\frac{\epsilon_{0}c^{2}}{2}\left(\nabla\times\hat{\boldsymbol{A}}\right)^{2}\right],
}
where $\epsilon_0$ is the vacuum permittivity, $c$ is the vacuum light speed,  $\hat{\boldsymbol{{A}}}(\boldsymbol{r})$ is the vector potential of photons, and  $\hat{\boldsymbol{\Pi}}(\boldsymbol{r})$ is its canonically conjugate variable. The integral is taken over a cubic whose volume is $V=L^3$ and we choose the Coulomb gauge 
\eqn{\nabla\cdot\hat{\boldsymbol{A}}=0.\label{CG}} 
 Assuming the periodic boundary conditions and introducing discrete wavenumbers $k_{i}=2\pi n_i/L$ with $i=x,y,z$ and $n_i$ being integer numbers, we represent the vector fields in terms of their Fourier components as
\eqn{\label{planeA}
\hat{\boldsymbol{A}}(\boldsymbol{r})&=&\sum_{\boldsymbol{k}\lambda}\sqrt{\frac{\hbar}{2\epsilon_{0}V\omega_{\boldsymbol{k}}}}\left(\hat{a}_{\boldsymbol{k}\lambda}e^{i\boldsymbol{k}\cdot\boldsymbol{r}}\boldsymbol{\epsilon}_{\boldsymbol{k}\lambda}+{\rm H.c.}\right),\\
\hat{\boldsymbol{\Pi}}(\boldsymbol{r})&=&-i\sum_{\boldsymbol{k}\lambda}\sqrt{\frac{\hbar\epsilon_{0}\omega_{\boldsymbol{k}}}{2V}}\left(\hat{a}_{\boldsymbol{k}\lambda}e^{i\boldsymbol{k}\cdot\boldsymbol{r}}\boldsymbol{\epsilon}_{\boldsymbol{k}\lambda}-{\rm H.c.}\right),
} 
where $\omega_{\boldsymbol{k}}=c\left|\boldsymbol{k}\right|$ is the vacuum photon dispersion, and $\hat{a}_{\boldsymbol{k}\lambda}$ ($\hat{a}_{\boldsymbol{k}\lambda}^\dagger$) is the annihilation (creation) operator of photons with momentum $\boldsymbol k$ and polarization $\lambda$. The Coulomb gauge~\eqref{CG} leads to the transversality condition of the orthonormal polarization vectors $\boldsymbol{\epsilon}_{\boldsymbol{k}\lambda}$:
\eqn{\label{polarization}
\boldsymbol{k}\cdot\boldsymbol{\epsilon}_{\boldsymbol{k}\nu}&=&0,\\
\boldsymbol{\epsilon}_{\boldsymbol{k}\lambda}^{\dagger}\boldsymbol{\epsilon}_{\boldsymbol{k}\nu}&=&\delta_{\lambda\nu}.\label{polarization2}
} 
The Hamiltonian can be diagonalized in the basis of momentum $\boldsymbol k$ and polarization $\lambda$ as 
\eqn{
\hat{H}_{{\rm light}}=\sum_{\boldsymbol{k}\lambda}\hbar\omega_{\boldsymbol{k}}\hat{a}_{\boldsymbol{k}\lambda}^{\dagger}\hat{a}_{\boldsymbol{k}\lambda},
}
where we omit the vacuum energy.

\subsubsection{Matter field}
We consider a dipolar insulator whose excitation is represented by a real-valued vector field $\hat{\boldsymbol\phi}(\boldsymbol{r})$. For instance, it can originate from infrared optical phonon modes corresponding to vibrations of ions around the equilibrium configuration in ionic crystals such as MgO, SiC and NaCl \cite{RR70}. In other setups, it can represent collective dipolar modes associated with assembled organic molecules, molecular crystals, and excitonic systems.
For the sake of simplicity, we neglect the bare dispersion of the dipolar mode, which is a common assumption in the analysis of optical phonons, and represent its frequency by $\Omega$. 
 The effective Hamiltonian of the matter field is thus given by
 \eqn{\hat{H}_{{\rm matter}}=\hat{H}_{0}+\hat{H}_{{\rm int}},\label{matterHam}}
 where $\hat{H}_0$ is the quadratic part 
 \eqn{\hat{H}_{0}=\int\frac{d^{3}r}{v}\left(\frac{\hat{\boldsymbol{\pi}}^{2}}{2M}+\frac{1}{2}M\Omega^{2}\hat{\boldsymbol{\phi}}^{2}\right),
 }
 and $\hat{H}_{\rm int}$ is the most relevant interaction
 \eqn{\label{phi4}
\hat{H}_{{\rm int}}=\int\frac{d^{3}r}{v}\,U\left(\hat{\boldsymbol{\phi}}\cdot\hat{\boldsymbol{\phi}}\right)^{2}.
 }
 Here, $\hat{\boldsymbol\pi}(\boldsymbol{r})$ is a conjugate field of the matter field $\hat{\boldsymbol{\phi}}(\boldsymbol{r})$, $v$ is the unit-cell volume in the insulator, $M$ is the effective mass, which corresponds to the reduced mass of two ions in the case of ionic crystals, and $U\geq 0$ characterizes the interaction strength. Physically, the interaction term can originate from, e.g., phonon anharmonicity. 
When we will discuss in Sec~\ref{sec4} the transition between the paraelectric phase with $\langle\boldsymbol{\phi}\rangle=\boldsymbol{0}$ and the ferroelectric phase with $\langle\boldsymbol{\phi}\rangle\neq\boldsymbol{0}$, the phonon frequency $\Omega^2$  should be considered as the tuning parameter governing the transition in the effective theory (cf. Fig.~\ref{fig1}).  
We note that the present model neglects the coupling between the gapped electron-hole excitations and soft phonons, which is typically considered as weak \cite{GDM72}.
  
The matter field $\hat{\boldsymbol\phi}(\boldsymbol{r})$ and its conjugate variable $\hat{\boldsymbol\pi}(\boldsymbol{r})$ can be quantized through the canonical quantization. In the Coulomb gauge, it is useful to decompose them in terms of the transverse and longitudinal components as $\hat{\boldsymbol{\phi}}(\boldsymbol{r})=\hat{\boldsymbol{\phi}}^{\parallel}(\boldsymbol{r})+\hat{\boldsymbol{\phi}}^{\perp}(\boldsymbol{r})$ and $\hat{\boldsymbol{\pi}}(\boldsymbol{r})=\hat{\boldsymbol{\pi}}^{\parallel}(\boldsymbol{r})+\hat{\boldsymbol{\pi}}^{\perp}(\boldsymbol{r})$. 
The quadratic part $\hat{H}_0$ of the matter Hamiltonian $\hat{H}_{\rm matter}$ can readily be diagonalized as (see Appendix~\ref{app_diag})
\eqn{\label{linmat}
\hat{H}_{0}=\hbar\Omega\left[\sum_{\boldsymbol{k}\lambda}\hat{\phi}_{\boldsymbol{k}\lambda}^{\perp\dagger}\hat{\phi}_{\boldsymbol{k}\lambda}^\perp+\sum_{\boldsymbol{k}}\hat{\phi}_{\boldsymbol{k}}^{\parallel\dagger}\hat{\phi}_{\boldsymbol{k}}^\parallel\right],
}
where $\hat{\phi}_{\boldsymbol{k}\lambda}^{\perp}$ ($\hat{\phi}_{\boldsymbol{k}\lambda}^{\perp \dagger}$) represents an annihilation (creation) operator of a transverse phonon with momentum $\boldsymbol k$ and polarization $\lambda$, while $\hat{\phi}_{\boldsymbol{k}}^{\parallel}$ ($\hat{\phi}_{\boldsymbol{k}}^{\parallel \dagger}$) is its longitudinal counterpart.

\subsubsection{Light-matter coupling}
The transverse dipolar matter field $\hat{\boldsymbol \phi}^\perp$ can directly couple to the electromagnetic field in a bilinear form. For instance, we recall that the field $\hat{\boldsymbol \phi}^\perp$ can represent transverse displacements of ionic charges with opposite signs and thus can couple to the electric field. In terms of the vector potential $\hat{\boldsymbol{A}}$, the light-matter interaction is thus given by
\eqn{
\hat{H}_{{\rm l-m}}=\int\frac{d^{3}r}{v}\left(-\frac{Z^*e}{M}\hat{\boldsymbol{\pi}}^\perp\cdot\hat{\boldsymbol{A}}+\frac{(Z^*e)^{2}}{2M}\hat{\boldsymbol{A}}^{2}\right)+V_{\rm C},\nonumber\\\label{lmint}
}
where $Z^*e$ is the effective charge of a dipolar mode, which corresponds to the charge of ions in the case of ionic crystals, and $V_{\rm C}$ is the Coulomb potential term whose explicit form is specified below. We note that, aside a constant factor, the conjugate field $\hat{\boldsymbol \pi}$ physically corresponds to the current operator $\boldsymbol{j}\propto\partial_t \boldsymbol{\phi}$ associated with the dipolar field $\hat{\boldsymbol \phi}$. 

We here invoke neither the dipole approximation nor the rotating wave approximation, which are often used in literature, but can break down especially in strong-coupling regimes \cite{LP15,LX18}. The light-matter interaction~\eqref{lmint} including both $\hat{\boldsymbol \pi}\cdot\hat{\boldsymbol A}$ and ${\boldsymbol A}^2$ terms allows one to explicitly retain the gauge invariance of the theory. It is also noteworthy that the light-matter coupling~\eqref{lmint} is nonvanishing even in the absence of photons. This interaction with the vacuum electromagnetic field originates from the zero-point fluctuations of the electric field and can be viewed as the light-matter interaction mediated by virtual photons. In this paper, we focus on such a low photon, quantum regime with no external pump.

Our choice of the Coulomb gauge leads to the nondynamical contribution $V_{\rm C}$ resulting from the constraint relating the longitudinal component of the electric field to that of the matter field. To see this explicitly, we first decompose the electric field in terms of the transverse and longitudinal components as
\eqn{
\hat{\boldsymbol{E}}=\hat{\boldsymbol{E}}^{\perp}+\hat{\boldsymbol{E}^{\parallel}},
}
which satisfy $\nabla\cdot\hat{\boldsymbol{E}}^{\perp}=0$ and $\nabla\times\hat{\boldsymbol{E}^{\parallel}}={\boldsymbol 0}$.
On one hand, the transverse part can be directly related to the vector potential as
\eqn{
\hat{\boldsymbol{E}}^{\perp}=-\frac{\partial\hat{\boldsymbol{A}}}{\partial t}.
}
On the other hand, the longitudinal part is subject to the constraint resulting from the Coulomb gauge
\eqn{\label{gaugeconst}
\hat{\boldsymbol{E}^{\parallel}}=-\frac{Z^*e}{\epsilon_{0}v}\hat{\boldsymbol{\phi}}^{\parallel},
}
where we recall that $\hat{\boldsymbol{\phi}}^{\parallel}$ is the longitudinal part of the matter field satisfying $\nabla\times\hat{\boldsymbol{\phi}}^{\parallel}=\boldsymbol{0}$. This longitudinal component leads to the expression of the Coulomb potential term $V_{\rm C}$ as follows \cite{Philbin_2010}:
\eqn{
V_{\rm C}=\frac{\epsilon_0}{2}\int d^3r\left(\hat{{\boldsymbol E}}^\parallel\right)^2=\frac{(Z^*e)^2}{2\epsilon_0 v}\int \frac{d^3r}{v}\left(\hat{{\boldsymbol \phi}}^\parallel\right)^2.
}

\subsubsection{Bulk dispersions}
One can diagonalize the quadratic part of the total Hamiltonian,
\eqn{\label{lmquad}
\hat{H}_{{\rm tot},0}=\hat{H}_{{\rm light}}+\hat{H}_{0}+\hat{H}_{{\rm l-m}},
}
which allows one to obtain an analytical expression of the bulk dispersions.
 The presence of light-matter interaction leads to the formation of the hybridized modes consisting of photons and phonons, known as the phonon polaritons. This hybridization strongly modifies the underlying dispersion relations such that the avoided crossing occurs between two polariton dispersions. At the quadratic level, we can decompose the quadratic Hamiltonian $\hat{H}_{{\rm tot},0}$ in Eq.~\eqref{lmquad} into two parts that solely include either transverse or longitudinal components of the field operators as follows:
 \eqn{
\hat{H}_{{\rm tot},0}=\hat{H}_{{\rm tot},0}^\perp+\hat{H}_{{\rm tot},0}^\parallel.
 }
  \begin{figure}[b]
\includegraphics[width=70mm]{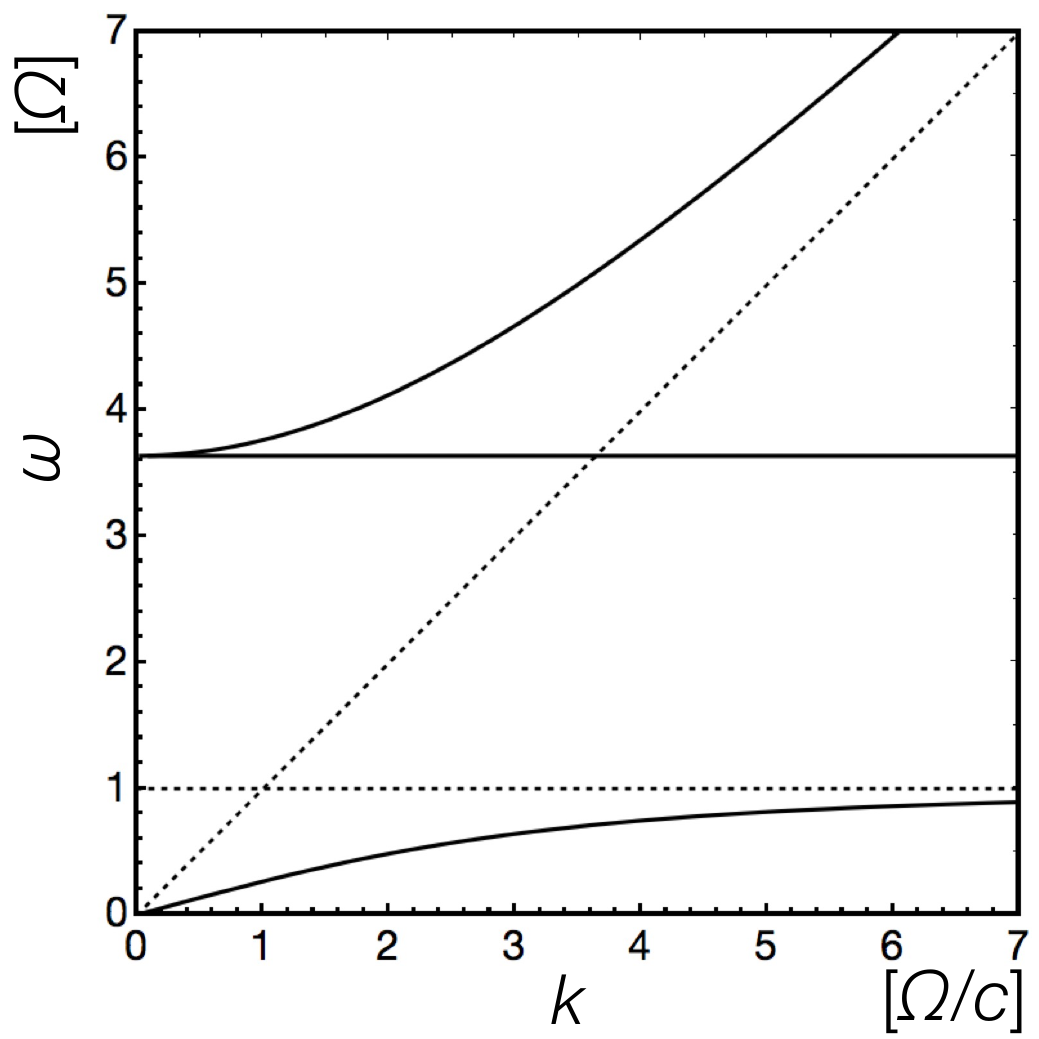} 
\caption{\label{fig4}
Bulk dispersions of the phonon polariton plotted against the norm of the three-dimensional wavevector $k=|\boldsymbol{k}|$. The solid curves represent the eigenvalues of the transverse modes (cf. Eq.~\eqref{eigvbulk}) for the upper ($s=+$) and the lower ($s=-$) phonon polaritons, which are obtained by diagonalizing the transverse part~$\hat{H}_{{\rm tot},0}^\perp$ of the quadratic light-matter Hamiltonian $\hat{H}_{{\rm tot},0}$. The solid horizontal line represents the nondispersive longitudinal phonon mode (cf. Eq.~\eqref{eigvbl}) obtained by diagonalizing the longitudinal part~$\hat{H}_{{\rm tot},0}^\parallel$. The dotted lines show the photon dispersion $\omega=ck$ and the nondispersive bare phonon frequency at $\omega=\Omega$. The parameter is chosen as $g=3.5\,\Omega$.  
}
\end{figure}
The dispersion relations of the transverse modes can now be obtained by diagonalizing $\hat{H}_{{\rm tot},0}^\perp$ (see Appendix~\ref{app_diag}).
The resulting Hamiltonian is
 \eqn{\label{totquadbulk}
\hat{H}_{{\rm tot},0}^\perp=\sum_{\boldsymbol{k}s\lambda}\hbar\omega_{\boldsymbol{k}s}\hat{\gamma}_{\boldsymbol{k}s\lambda}^{\perp\dagger}\hat{\gamma}_{\boldsymbol{k}s\lambda}^\perp,
 }
 where $s=\pm$ denotes an upper or a lower bulk dispersion with eigenfrequencies
 \eqn{\label{eigvbulk}
\omega_{\boldsymbol{k}\pm}=\sqrt{\frac{\omega_{\boldsymbol{k}}^{2}+\Omega^{2}+g^{2}\pm\sqrt{\left(\omega_{\boldsymbol{k}}^{2}+\Omega^{2}+g^{2}\right)^{2}-4\omega_{\boldsymbol{k}}^{2}\Omega^{2}}}{2}},\nonumber\\
 }
 and $\hat{\gamma}_{\boldsymbol{k}s\lambda}^\perp$ ($\hat{\gamma}_{\boldsymbol{k}s\lambda}^{\perp \dagger}$) represents an annihilation (creation) operator of a transverse phonon polariton  with momentum $\boldsymbol k$, polarization $\lambda$, and mode $s$. We here introduce the coupling strength $g$ as
\eqn{
g=\sqrt{\frac{(Z^*e)^{2}}{\epsilon_{0}Mv}}.
}
The solid curves in Fig.~\ref{fig4} show the two branches of the bulk phonon polaritons. The splitting between the upper and lower dispersive modes is characterized by the coupling strength $g$. The upper branch exhibits a quadratic dispersion at low $k=|\boldsymbol k|$ and is gapped, i.e., it takes a finite value $\sqrt{\Omega^2+g^2}$ at $k=0$,  while the lower one linearly grows from zero and saturates at $\Omega$ even in the limit of ${ k}\to\infty$. 

Meanwhile, the longitudinal part of the total quadratic Hamiltonian $\hat{H}_{{\rm tot},0}^\parallel$ only contains the longitudinal matter field $\hat{\boldsymbol \phi}^\parallel$ and can be readily diagonalized as
\eqn{\label{eigvbl}
\hat{H}_{{\rm tot},0}^\parallel=\hbar\Omega^{\parallel}\sum_{\boldsymbol k}\hat{ \phi}^{\parallel\dagger}_{{\boldsymbol k}}\hat{ \phi}_{\boldsymbol k}^{\parallel},
}
where $\Omega^\parallel=\sqrt{\Omega^2+g^2}$ is the longitudinal phonon frequency \cite{MCP10}. We note that the longitudinal mode is independent of the momentum $\boldsymbol k$ and is thus nondispersive as shown in the solid horizontal line in Fig.~\ref{fig4}.

\subsection{Cavity configuration}
\begin{figure}[b]
\includegraphics[width=86mm]{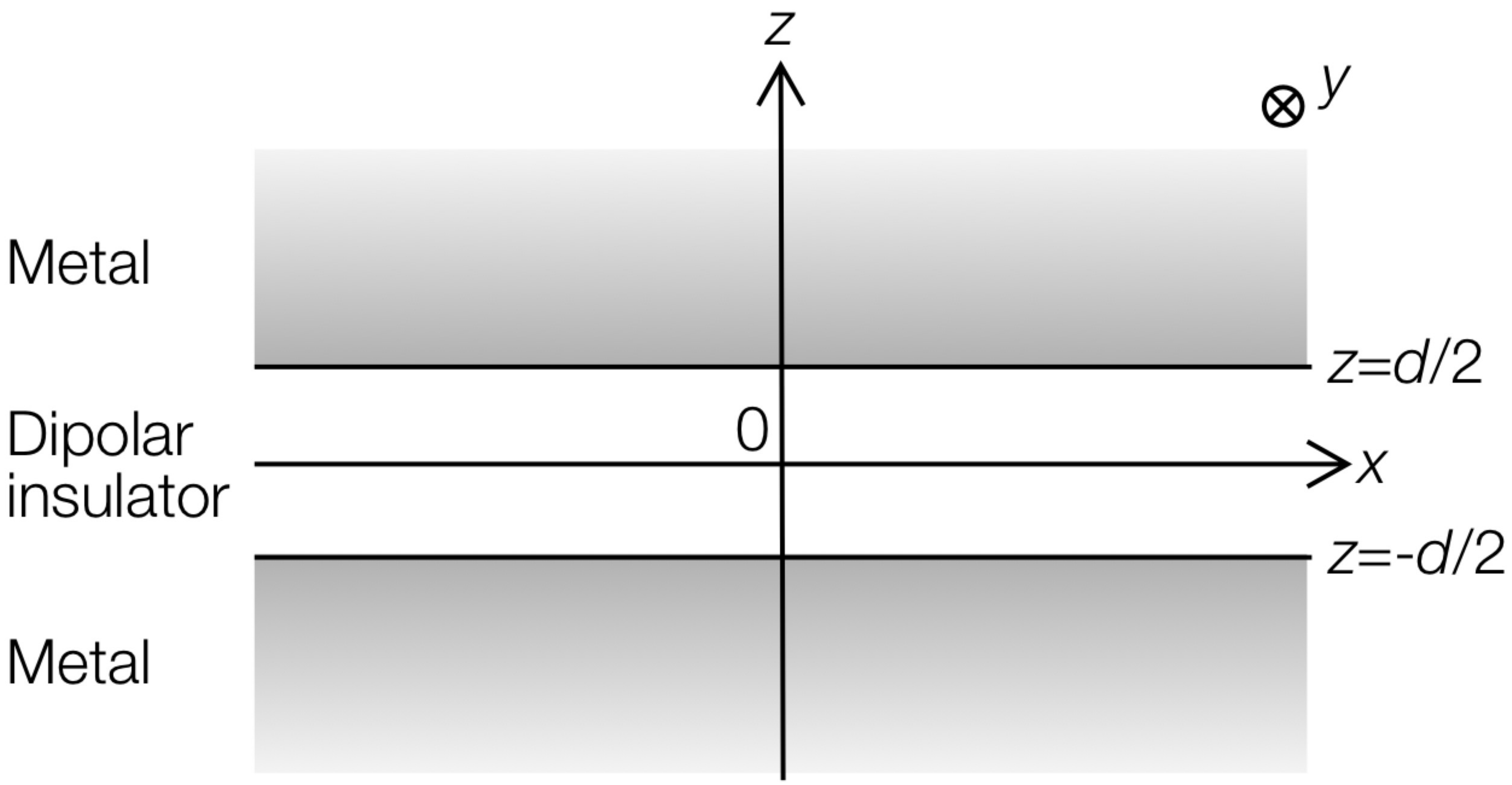} 
\caption{\label{fig5}
Schematic figure illustrating the cavity configuration considered in this paper. A dipolar insulator whose center is positioned at $z=0$ is sandwiched between two semi-infinite metals with interfaces at $z=\pm d/2$.  
}
\end{figure}
We next consider the cavity setting and take into account confinement effects as well as the hybridization of electromagnetic fields with plasmons in metal mirrors in addition to collective dipolar modes.
We focus on the prototypical heterostructure configuration consisting of a dipolar insulator sandwiched between two semi-infinite ideal metals (see Fig.~\ref{fig5}). We consider eigenmodes propagating with the two-dimensional in-plane momentum $\boldsymbol q$ while the formalism developed in this section can be extended to deal with more complex setups such as  fabricated metal surface structures. 

Thin-film configurations considered here have been previously applied to light-emitting devices \cite{JV00,KDM08}, the electron-tunneling emission \cite{LJ76}, and the transmitting waveguide \cite{SA00}.
 One of the main novelties introduced by this paper is to reveal the great potential of heterostructure configurations as a quantum electrodynamic setting towards controlling collective matter excitations and, in particular, to point out a possibility of inducing the superradiant-type quantum phase transition via the vacuum electromagnetic environment. More specifically, analyzing the hybridization of cavity electromagnetic fields and matter excitations, we  demonstrate that the present cavity architecture  enriches the underlying dispersion relations in a way qualitatively different from the corresponding bulk case. Including phonon nonlinearities, we will further show that the heterostructure  configuration enables one to significantly soften the  phonon modes, which ultimately induces the structural phase transition.

\subsubsection{Hamiltonian}
To reveal essential features of the present cavity setting, we consider an 
 ideal metal with the Drude property whose plasma frequency is denoted as $\omega_{p}=\sqrt{n_e e^2/(m_e\epsilon_0)}$, where $e$ is the elementary electric charge, $n_e$ is the electron density in the metal, and $m_e$ is the electronic mass. We assume that all the materials are nonmagnetic such that the relative magnetic permeability $\mu$ is always taken to be $\mu=1$.  As illustrated in Fig.~\ref{fig5}, we consider a setup with the center of a dipolar insulator being positioned at $z=0$ while the insulator-metal interfaces being positioned at $z=\pm d/2$. We again choose the Coulomb gauge $\nabla\cdot\hat{\boldsymbol A}=0$ and thus the Hamiltonian $\hat{H}_{\rm light}$ of the free electromagnetic field  is the same as in the bulk case (cf. Eq.~\eqref{emfree}) while the matter Hamiltonian is modified as
 \eqn{
 \hat{H}_{{\rm matter}}&=&\hat{H}_{0}+\hat{H}_{{\rm int}},\label{mimmat}\\
 \hat{H}_{0}&=&\int_{{\rm I}}\frac{d^{3}r}{v}\left(\frac{\hat{\boldsymbol{\pi}}^{2}}{2M}+\frac{1}{2}M\Omega^{2}\hat{\boldsymbol{\phi}}^{2}\right),\label{mimquad}\\
\hat{H}_{{\rm int}}&=&\int_{{\rm I}}\frac{d^{3}r}{v}\,U\left(\hat{\boldsymbol{\phi}}\cdot\hat{\boldsymbol{\phi}}\right)^{2},\label{mimint}
 }
 where we define the integral over the insulator region as $\int_{{\rm I}}d^{3}r\equiv\int d^{2}r\int_{-d/2}^{d/2}dz$.
 The light-matter coupling term is also modified as 
 \eqn{
\hat{H}_{{\rm l-m}}\!&=&\!\int_{{\rm M}}d^{3}r\,\frac{\epsilon_{0}\omega_{p}^{2}}{2}\hat{\boldsymbol{A}}^{2}\nonumber\\
&+&\!\int_{{\rm I}}\frac{d^{3}r}{v}\!\left(\!-\frac{Z^*e}{M}\hat{\boldsymbol{\pi}}^\perp\cdot\hat{\boldsymbol{A}}\!+\!\frac{(Z^*e)^{2}}{2M}\hat{\boldsymbol{A}}^{2}\right)\!+\!V_{\rm C}, \label{mimlm}
 }
 where the first term describes the plasmon coupling in the metal regions defined as  
$\int_{{\rm M}}d^{3}r\equiv\int d^{2}r\left(\int_{-\infty}^{-d/2}+\int_{d/2}^{\infty}\right)dz$, and the second term is the coupling between the insulator and the cavity electromagnetic field. We recall that the Coulomb gauge leads to the additional constraint on the longitudinal components of the electric and matter fields in the insulator region $|z|<d/2$: 
\eqn{\label{gaugeconst2}
\hat{\boldsymbol{E}^{\parallel}}=-\frac{Z^*e}{\epsilon_{0}v}\hat{\boldsymbol{\phi}}^{\parallel}.
}
As in the bulk case, this constraint results in the Coulomb potential term $V_{\rm C}$ originating from the longitudinal dipolar modes as follows:
\eqn{
V_{\rm C}=\frac{\epsilon_0}{2}\int_{\rm I} d^3r\left(\hat{{\boldsymbol E}}^\parallel\right)^2=\frac{(Z^*e)^2}{2\epsilon_0 v}\int_{\rm I} \frac{d^3r}{v}\left(\hat{{\boldsymbol \phi}}^\parallel\right)^2.
}

Summarizing, the present cavity light-matter system consists of two canonically conjugate pairs of variables $(\hat{{\boldsymbol \Pi}},\hat{{\boldsymbol A}})$  and $(\hat{{\boldsymbol \pi}},\hat{{\boldsymbol \phi}})$; the former describes the dynamical electromagnetic degrees of freedom and only have transverse components while the latter describes matter excitations and have both transverse and longitudinal parts. Their time evolution is governed by the Hamiltonian consisting of Eqs.~\eqref{emfree}, \eqref{mimmat} and \eqref{mimlm}, and is subject to the constraint~\eqref{gaugeconst2}.

We note that the use of the Coulomb gauge remains to be a convenient choice even in the present case of the inhomogeneous geometry. An alternative naive choice would be the condition $\nabla\cdot[\epsilon({\boldsymbol r},\omega)\hat{{\boldsymbol A}}({\boldsymbol r},\omega)]=0$ with $\epsilon$ being the relative permittivity in the medium, which might be viewed as a quantum counterpart of the gauge constraint in the macroscopic Maxwell equations. However, it has been pointed out that in inhomogeneous media this gauge choice suffers from the difficulty of performing the canonical quantization of the variables due to the frequency-domain formulation of the gauge constraint \cite{Philbin_2010}. In contrast, the Coulomb gauge chosen here can be still imposed on a general inhomogeneous system without such difficulties; we thus employ it throughout this paper.  

\subsubsection{Elementary excitations}

\begin{figure*}[t]
\includegraphics[width=172mm]{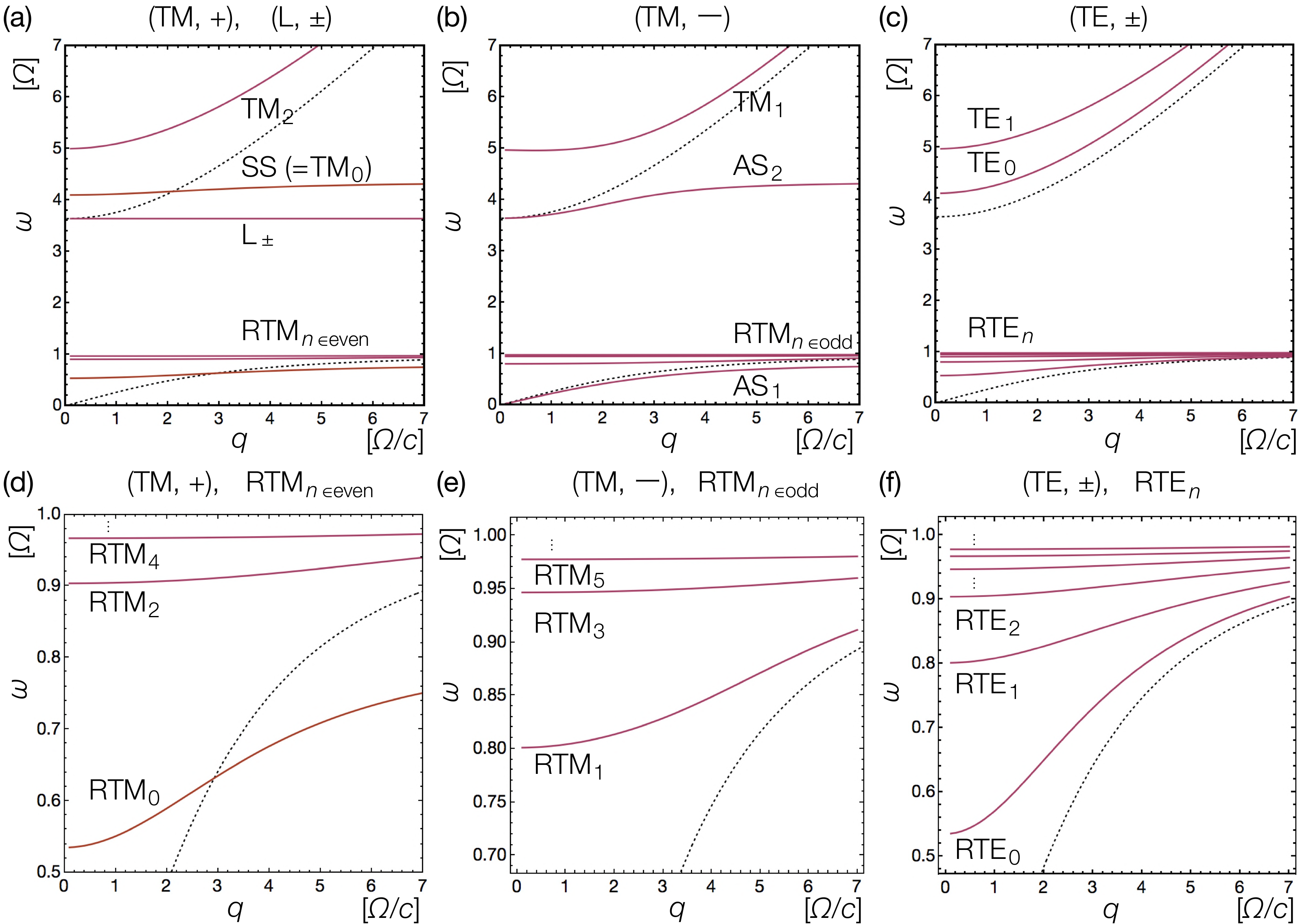} 
\caption{\label{fig6}
Elementary excitations of the cavity setting in different sectors $(\Lambda,P)$ with $\Lambda=$TM, TE, L denoting the transverse magnetic (TM), transverse electric (TE), or longitudinal (L) polarization and $P=\pm$ being the parity of transverse components of the electric field against the inversion of the direction perpendicular to the plane. Black dotted curves indicate the bulk dispersions.
The top panels plot excitation energies $\omega$ against in-plane momentum $q$ in the sectors (a) (TM,$+$) and (L,$\pm$), (b) (TM,$-$), and (c) (TE,$\pm$), respectively. The bottom panels (d), (e), and (f) show the corresponding magnified views below the phonon frequency $\Omega$. In (a) and (b), the TM$_{1,2}$ modes represent the two highest modes in the TM sector, the SS (=TM$_0$) mode represents the symmetric surface (SS) mode, the AS$_{1,2}$ modes represent the antisymmetric surface (AS) modes whose dipolar moments point out of the plane at low $q$. The longitudinal L$_{\pm}$ modes have the same nondispersive energy independent of the momentum $q$ and the parity $P=\pm$. The assignment of the label TM$_0$ to the SS mode is in accordance with the convention used in studies of plasmonics \cite{SH12}.  In (d) and (e), the RTM$_{n}$ modes with $n=0,1,2\ldots$ represent the resonant TM modes close to the dipolar-phonon frequency $\Omega$. The RTM$_{n\geq 1}$ modes extend over the insulator region while the most-softened mode RTM$_0$ is essentially the surface mode localized at the interfaces analogous to the SS mode. In (c), the TE$_{0,1}$ modes represent the two highest modes in the TE sector. In (f), the RTE$_n$ modes with $n=0,1,2\ldots$ represent the resonant TE modes; all of them extend over the insulator region.
The parameters are $d=c/\Omega$, $g=3.5\,\Omega$, and $\omega_p=5\,\Omega$. For the sake of visibility, the plasma frequency is set to be a rather low value while its specific choice will not qualitatively affect the thermodynamic phase.
}
\end{figure*}

{\renewcommand{\arraystretch}{1.2}
\begin{table*}
\caption{Summary of physical properties for each of elementary excitations. The labels for each band in the first column are consistent with those indicated in Fig.~\ref{fig6}.  The second and third columns  represent the parity $P=\pm$  and the polarization  $\Lambda\in\{{\rm TM},{\rm TE},{\rm L}\}$ of each mode, respectively. The fourth column indicates the direction of the dipole moments in the insulator at low in-plane wavevector $q$ with respect to the plane parallel to the interfaces. The fifth column shows whether or not the divergence of the electric field is vanishing. The sixth and seventh columns represent whether the longitudinal wavenumber $\kappa_\eta$ takes a real value or a purely imaginary value at low and high $q$, respectively. The subscript $\eta\in\{{\boldsymbol q},n,\lambda\}$ with $\lambda=(\Lambda,P)$ represents a set of all the variables specifying each eigenmode. The final two columns indicate the physical nature of each hybridized band at low and high $q$, where we abbreviate elementary excitations as plasmons (PL), photons (PT), phonons (PN), polaritons (P), and specify surface modes by S.  }\label{TETMtable}
\begin{tabular}{@{\hspace{0mm}} l @{\hspace{9.5mm}} c @{\hspace{9mm}}c @{\hspace{9mm}} c @{\hspace{9mm}}c @{\hspace{9mm}} c @{\hspace{8mm}} c @{\hspace{6mm}} c @{\hspace{8mm}} c @{\hspace{1mm}}}
\hline
\hline
\multirow{2}{*}{Band} & \multirow{2}{*}{$P$} &\multirow{2}{*}{$\Lambda$}  & \multirow{2}{*}{Dipole at low ${q}$}  & \multirow{2}{*}{Divergence} & \multicolumn{2}{c}{$\kappa_{\eta}$} &  \multicolumn{2}{c}{Mode} \\
\cmidrule(r){6-7} \cmidrule(r){8-9}
&&&&& low $q$ & high $q$ & low $q$ & high $q$\\
\hline
TM$_{2}$ & $+$ & TM & In-plane & zero & Real & Real & PL & PT \\
TM$_{1}$ & $-$ & TM & In-plane & zero & Real & Real & PL-P & PT \\
RTM$_{n\geq 1}$ &  sgn($n$) & TM & In-plane & zero & Real & Real & PN-P & PN \\
RTM$_{0}$ & $+$ & TM & In-plane & zero & Real & Imaginary & PN-P & S-PN-PL-P \\
L$_\pm$ & $\pm$ & L & Out-of-plane & nonzero & Real & Real & PN-P & PN-P \\
SS ($=$TM$_{0}$) & $+$ & TM & In-plane & zero & Real & Imaginary & PN-P & S-PN-PL-P \\
AS$_{2}$ & $-$ & TM & Out-of-plane & zero & Imaginary & Imaginary & PN-P & S-PN-PL-P \\
AS$_{1}$ & $-$ & TM & Out-of-plane & zero & Imaginary & Imaginary & PT & S-PN-PL-P \\
TE$_{1}$ & $-$ & TE & In-plane & zero & Real & Real & PL & PT \\
TE$_{0}$ & $+$ & TE & In-plane & zero & Real & Real & PL-P & PT \\
RTE$_{n\geq 0}$ &  sgn($n$) & TE & In-plane & zero & Real & Real & PN-P & PN\\
\hline
\hline
\end{tabular}
\end{table*}

The quadratic part of the total cavity Hamiltonian can still be diagonalized, which allows us to identify the hybridized eigenmodes and the corresponding elementary excitation energies. To this end, we first expand the electromagnetic vector potential in terms of the transverse elementary modes. Owing to the spatial invariance on the $xy$ plane, we obtain
\eqn{
\hat{\boldsymbol{A}}(\boldsymbol{r})\!&=&\!\sum_{\boldsymbol{q}n\lambda}\!\sqrt{\frac{\hbar}{2\epsilon_{0}A\omega_{\boldsymbol{q}n\lambda}}}\!\left(\hat{a}_{\boldsymbol{q}n\lambda}\boldsymbol{U}_{\boldsymbol{q}n\lambda}^{\perp}(\boldsymbol{r})\!+\!{\rm H.c.}\right)\!,\\
\boldsymbol{U}_{\boldsymbol{q}n\lambda}^{\perp}(\boldsymbol{r})&=&e^{i\boldsymbol{q}\cdot\boldsymbol{\rho}}\boldsymbol{u}^\perp_{\boldsymbol{q}n\lambda}(z),
}
where $A=L^2$ is the area of interest on the $xy$ plane, $\boldsymbol q$ is the two-dimensional in-plane wavevector $\boldsymbol{q}=(q_{x},q_{y})^{{\rm T}}$, and $\boldsymbol{\rho}=(x,y)^{{\rm T}}$ is the position vector on the $xy$ plane. The transverse mode function $\boldsymbol{U}_{\boldsymbol{q}n\lambda}^{\perp}$ satisfies $\nabla\cdot\boldsymbol{U}_{\boldsymbol{q}n\lambda}^{\perp}=0$.

Because the spatial invariance along the $z$ direction is lost in the  heterostructure configuration, an elementary excitation is now labeled by a discrete number $n\in\mathbb{N}$ (instead of $k_z$ in the bulk case) as well as the two-dimensional vector $\boldsymbol q$. We also introduce a discrete variable $\lambda=(\Lambda,P)$ that includes the polarization label $\Lambda\in\left\{ {\rm TM},{\rm TE}\right\}$ and the parity label $P=\pm$. Here, the polarization label $\Lambda$ in the discrete subscript $\lambda$ indicates either the transverse magnetic (TM) or electric (TE) polarization associated with the zero magnetic or electric field in the $z$ direction:
 \eqn{
\left(\boldsymbol{u}^\perp_{\boldsymbol{q}n\lambda}(z)\right)_{z}&=&0\;\;{\rm for}\;\Lambda={\rm TM},\\
\left(\nabla\times\boldsymbol{U}_{\boldsymbol{q}n\lambda}^{\perp}(\boldsymbol{r})\right)_{z}&=&0\;\;{\rm for}\;\Lambda={\rm TE}.
 }
 Meanwhile, the parity label $P$ indicates the symmetry of the mode function with respect to the spatial inversion along the $z$ direction. 
From the parity symmetry of the Maxwell equations, it follows that the $z$ component of the mode function must have the opposite parity to that of the $x,y$ components (see Appendix~\ref{app2}). We thus assign the parity label $P$ to each mode function according to 
\eqn{\label{upar1}
\left(\boldsymbol{u}^\perp_{\boldsymbol{q}n\lambda}(-z)\right)_{x,y}&=&P\left(\boldsymbol{u}^\perp_{\boldsymbol{q}n\lambda}(z)\right)_{x,y},\\
\left(\boldsymbol{u}^\perp_{\boldsymbol{q}n\lambda}(-z)\right)_{z}&=&-P\left(\boldsymbol{u}^\perp_{\boldsymbol{q}n\lambda}(z)\right)_{z}.\label{upar2}
}
Finally, we can impose the orthonormal conditions on the mode functions as
\eqn{
\int dz\,\boldsymbol{u}^{\perp\dagger}_{\boldsymbol{q}n\lambda}\boldsymbol{u}^\perp_{\boldsymbol{q}m\nu}=\delta_{nm}\delta_{\lambda\nu}.
}

We can expand the matter degrees of freedom in the similar manner as in the electromagnetic field. Because the matter field resides only in the insulator region, it is natural to expand the transverse components of the dipolar matter field as
\eqn{\hat{\boldsymbol{\phi}}^{\perp}(\boldsymbol{r})\!=\!\sqrt{\frac{\hbar}{2MN_{d}\Omega}}\sum_{\boldsymbol{q}n\lambda}\left(\hat{\phi}_{\boldsymbol{q}n\lambda}^{\perp}\boldsymbol{f}_{\boldsymbol{q}n\lambda}^{\perp}(z)e^{i\boldsymbol{q}\cdot\boldsymbol{\rho}}+{\rm H.c.}\right),\nonumber\\}
where $N_{d}=A/v=N/d$ is the number of modes per thickness, and $\{\boldsymbol{f}_{\boldsymbol{q}n\lambda}^{\perp}\}$ is a complete set of transverse orthonormal functions in the insulator region satisfying $\nabla\cdot(\boldsymbol{f}_{\boldsymbol{q}n\lambda}^{\perp}e^{i\boldsymbol{q}\cdot\boldsymbol{\rho}})=0$. We note that a position variable $\boldsymbol r$ in the matter field $\hat{\boldsymbol \phi}$ should be interpreted as the parameters restricted to the region $|z|<d/2$. In the discrete variable $\lambda=(\Lambda,P)$, the polarization label takes either TM or TE modes, $\Lambda\in\{{\rm TM},{\rm TE}\}$, and the parity label $P=\pm$ is defined in the same manner as in Eqs.~\eqref{upar1} and \eqref{upar2}.  
Another possible matter excitations are the longitudinal modes, for which we can expand the matter field as
\eqn{
\hat{\boldsymbol{\phi}}^{\parallel}(\boldsymbol{r})=\sqrt{\frac{\hbar}{2MN_{d}\Omega}}\sum_{\boldsymbol{q}n\lambda}\left(\hat{\phi}_{\boldsymbol{q}n\lambda}^{\parallel}\boldsymbol{f}_{\boldsymbol{q}n\lambda}^{\parallel}(z)e^{i\boldsymbol{q}\cdot\boldsymbol{\rho}}+{\rm H.c.}\right),\nonumber\\
}
where $\{\boldsymbol{f}_{\boldsymbol{q}n\lambda}^{\parallel}\}$ is a complete set of longitudinal orthonormal functions in the insulator region, which satisfies $\nabla\times (\boldsymbol{f}_{\boldsymbol{q}n\lambda}^{\parallel}e^{i\boldsymbol{q}\cdot\boldsymbol{\rho}})={\boldsymbol 0}$. We denote this mode by the label $\lambda=({\rm L},P)$, where the polarization label $\Lambda$ is fixed to be the longitudinal one $\Lambda={\rm L}$ while the parity $P=\pm$ of the mode function is defined in the same way as in the transverse case. The longitudinal mode associates with an excitation of the longitudinal electric field via Eq.~\eqref{gaugeconst2}.

An explicit form of the transverse mode functions $\{\boldsymbol{u}^\perp_{\boldsymbol{q}n\lambda}\}$ and $\{\boldsymbol{f}^\perp_{\boldsymbol{q}n\lambda}\}$, the longitudinal mode functions $\{\boldsymbol{f}^\parallel_{\boldsymbol{q}n\lambda}\}$, and the corresponding excitation energies $\{\omega_{\boldsymbol{q}n\lambda}\}$ can be determined from solving the eigenvalue problem with imposing suitable boundary conditions at the interfaces as detailed in Appendix~\ref{app2}. 
In the transverse modes, we note that both matter and light fields as well as the hybridized eigenmodes share the same spatial profile represented by $\boldsymbol{u}^\perp_{\boldsymbol{q}n\lambda}$ owing to the bilinear form of the light-matter coupling. Meanwhile, the longitudinal modes consist of only the matter field and their spatial profiles are characterized in terms of the mode functions $\boldsymbol{f}^\parallel_{\boldsymbol{q}n\lambda}$.  
It is also noteworthy that an excitation energy $\omega_{\boldsymbol{q}n\lambda}$ of an elementary eigenmode now explicitly depends on the polarization $\Lambda\in\{{\rm TM},{\rm TE},{\rm L}\}$ in contrast to the bulk case in Eq.~\eqref{eigvbulk}. 
In terms of the obtained elementary eigenmodes, the quadratic part of the total light-matter Hamiltonian,  
$\hat{H}_{{\rm tot},0}=\hat{H}_{{\rm light}}+\hat{H}_{0}+\hat{H}_{{\rm l-m}},$ 
can be diagonalized as 
\eqn{
\hat{H}_{{\rm tot},0}=\sum_{\boldsymbol{q}n\lambda}\hbar\omega_{\boldsymbol{q}n\lambda}\hat{\gamma}_{\boldsymbol{q}n\lambda}^{\dagger}\hat{\gamma}_{\boldsymbol{q}n\lambda},
}
where $\hat{\gamma}_{\boldsymbol{q}n\lambda}$ ($\hat{\gamma}_{\boldsymbol{q}n\lambda}^\dagger$) represents an annihilation (creation) operator of a hybridized elementary excitation with in-plane wavevector $\boldsymbol q$ and discrete labels $n$ and $\lambda$. The spatial profile of each excitation is characterized by an eigenmode function ${\boldsymbol u}^\perp_{{\boldsymbol q}n\lambda}$ or $\boldsymbol{f}_{\boldsymbol{q}n\lambda}^{\parallel}$ depending on the polarization, whose explicit functional forms are given in Appendix~\ref{app2}.

Figure~\ref{fig6} shows the elementary excitations in the cavity setting for each sector of $\lambda=(\Lambda,P)$, and Table~\ref{TETMtable} summarizes the corresponding physical properties. Figures~\ref{fig6}(a) and (b) plot the in-plane dispersions in the sectors $\lambda=({\rm TM},+)$ and $({\rm L},\pm)$, and $\lambda=({\rm TM},-)$, respectively. Their magnified plots around the phonon frequency $\Omega$ are also given in Figs.~\ref{fig6}(d) and (e). These rich band structures in the TM polarization have  different physical origins as described below.

First of all, a salient feature of the obtained cavity polariton is the emergence of a large number of soft-phonon modes lying below the bare phonon frequency $\Omega$, which we term as the resonant TM (RTM) modes  (see Fig.~\ref{fig6}(d,e)). The RTM modes are labeled by an integer number $n=0,1,2\ldots$ as RTM$_n$. The  parity $P$ of the corresponding mode function ${\boldsymbol u}_{{\boldsymbol q}n\lambda}^\perp$ coincides with the parity of $n$. The RTM$_n$ modes with $n\geq 1$ lie between $\Omega$ and the lower bulk dispersion (plotted as the lower black dotted curve in Fig.~\ref{fig6}), and share the common structures. For instance, all of the RTM$_{n\geq 1}$ modes start from frequencies slightly below $\Omega$ and saturate to it at high $q\equiv|\boldsymbol{q}|$. Thus, these modes predominantly consist of the  hybridization of dipolar phonons and cavity electromagnetic fields at low $q$ while the phonon contribution eventually becomes dominant at high $q$. The field amplitudes of these RTM$_{n\geq 1}$ modes extend over the insulator region owing to the real-valuedness of the longitudinal wavenumber $\kappa_\eta$ over the entire plane of the in-plane wavevector $\boldsymbol q$.  Here, the subscript $\eta$ is used to summarize all the indices as $\eta=\{{\boldsymbol q},n,\lambda\}$.  The origin of the emergence of many RTM modes below the bare phonon frequency can be understood from simple physical arguments based on the frequency-dependent dielectric functions as discussed in Sec.~\ref{sec_resori}.

Remarkably, the lowest resonant mode RTM$_0$, which is the most-softened phonon mode among all the elementary excitations in all the sectors, has a qualitatively different physical nature compared to the RTM$_{n\geq 1}$ modes discussed above. Its longitudinal wavenumber $\kappa_\eta$ changes from a real value to a purely imaginary one as the dispersion crosses the lower bulk dispersion in the bulk (see Fig.~\ref{fig6}(d)). Moreover, the saturated frequency value of the RTM$_0$ mode at $q\to\infty$ remains substantially below $\Omega$ with a nonvanishing gap.   Accordingly, the RTM$_0$ mode can be interpreted as the low-energy analogue of the surface polariton mode labeled as the SS mode in Fig~\ref{fig6}(a); both modes result from the intrinsic hybridization of dipolar phonons, cavity electromagnetic fields, and plasmons while the major contribution at high $q$ comes from the phonon (plasmon) part in the case of the RTM$_0$ (SS) mode.

We now proceed to  discuss the symmetric and antisymmetric surface modes lying in the frequency regime $\sqrt{\Omega^2+g^2}<\omega<\omega_p$, which are denoted as SS(=TM$_0$) and AS$_2$ in Figs.~\ref{fig6}(a) and (b), respectively. They essentially arise from the splitting of the coupled surface phonon polaritons at two metal-insulator interfaces \cite{EEN69}.  
 These surface modes result from the intrinsic hybridization of the collective dipolar mode, electromagnetic fields in the cavity, and plasmons in metal mirrors, and can emerge only in the case of the TM polarization, in which the continuity of the $z$ component of the electric field can be mitigated. Thus, the surface modes including the SS(=TM$_0$), AS$_{1,2}$, and RTM$_0$ modes can be excited only by $p$-polarized light.
 Additional discussion of the hybridized eigenmodes in our cavity system can be found in Appendix~\ref{app_hyb}.

Several remarks are in order. Firstly, we here emphasize the importance of explicitly including plasmon contributions into the analysis. This treatment ensures that the tails of the evanescent fields properly extend into the surrounding metal mirrors. In many of the previous studies, the mirrors are assumed to be  perfectly reflective such that the electric fields must exactly vanish at the interfaces and are absent in the metal regions. In such analyses, the physical nature of the localized surface modes discussed here cannot be addressed appropriately. For instance, the most-softened phonon mode RTM$_{0}$ is one important example of such localized modes; an analysis assuming the perfect reflectivity cannot capture its unique features, including the qualitative changes associated with increasing in-plane momentum $q$ (cf.~Fig.~\ref{fig7} in Appendix~\ref{app_hyb}), the nonvanishing softening from the bare phonon frequency $\Omega$ even in the limit $q\to\infty$ (cf.~Fig.~\ref{fig6}(d)), as well as the emergence of roton-type excitations (cf.~Fig.~\ref{fig8} in Appendix~\ref{app_hyb}). Importantly, the correct treatment of this RTM$_0$ mode turns out to be crucial especially when we discuss the superradiant-type quantum phase transition as detailed in the next section. Meanwhile, we remark that the reflectivity does not qualitatively change physical properties of the TE modes, which are analogous to the modes supported in Fabry-Perot cavities \cite{PP98}.

Secondly, we remark that the rich structures of the elementary excitations demonstrated above can further be  tuned by varying the thickness $d$ of the dipolar insulator. The increase of $d$ generally leads to a generation of more delocalized bulk modes extending over the insulator. Thus, aside from the localized surface modes, the eigenmodes become denser and eventually converge to the corresponding bulk dispersions in the limit of $d\to\infty$. In contrast, the decrease of $d$ leads to the filtering of the delocalized modes; for instance, a degenerate frequency of the TE$_0$ and SS modes at zero-in-plane momentum shifts up for a smaller $d$. Thus, with decreasing $d$, the localized nature of the modes plays more and more crucial roles in determining the physical properties of the present heterostructure configuration. We assess the effects of changing the thickness in more detail  in the next section by including the most relevant nonlinear effect in a self-consistent manner. The nonlinearity renormalizes the value of the bare phonon frequency $\Omega$ and thus effectively modifies the underlying elementary excitations. We will see that a thinner insulator is more preferable for realizing the cavity-enhanced ferroelectricity. The reason is that a thin heterostructure configuration can alleviate the adverse effect of the  interaction among the energetically dense phonon excitations and is thus advantageous for inducing the ultimate softening of the RTM$_0$ mode, which causes the ferroelectric instability. 

Finally, we comment on the importance of considering all the above-mentioned   hybridized modes in the cavity geometry, thus keeping track of the continuum of excitations for every dispersion. 
In general, nonresonant modes can give rise to substantial modifications of dipolar moments in the confined geometry as compared to their counterparts in free space, especially in the regime of ultrastrong coupling. This phenomenon has been discussed for systems with dipole-dipole interaction and demonstrated to have a qualitative effect on thermodynamic properties of a system \cite{PP20,SM20}; we here recall that dipolar interactions can be understood as arising from the adiabatic elimination of nonresonant modes under certain conditions \cite{DB18}. In our analysis, all the confinement effects due to the cavity, including dipole-dipole interactions, are consistently included into the theory without relying on approximations such as rotating-wave approximations or adiabatic eliminations, which can fail in the presence of closely degenerate low-energy modes such as the RTM/RTE modes as discussed here. 
The inclusion of all the continuum modes becomes further important when we will study the phonon nonlinearity below. There, due to mode couplings mediated by the nonlinearity, an energy of individual modes can depend on fluctuations of  the other modes.

\subsubsection{Origin of the resonant soft-phonon modes\label{sec_resori}}
We here provide simple explanations to elucidate the origin of the resonant softened phonon modes that are denoted by the labels RTM and RTE in Fig.~\ref{fig6}. To this end, we consider a frequency regime in which the metallic permittivity is negative $\epsilon_p(\omega)<0$ while the insulator dielectric function is positive $\xi(\omega)>0$ (see also Eqs.~\eqref{metdiele} and \eqref{insdiele} in App.~\ref{app2}). To be specific, we focus on the RTM modes at the zero in-plane momentum $q=0$ while similar arguments can be given for the corresponding RTE modes as well. 

Inside the insulator region, the RTM modes with $q=0$ have the spatial profiles of the magnetic field either $\sin(\kappa z)$ or $\cos(\kappa z)$, which we denote by the labels $P=+$ or $P=-$, respectively. We here recall that the parity of the eigenmodes is defined by that of the transverse components of the electric field. In these cases, the longitudinal wavevector $\kappa$ should satisfy the following equations:
\eqn{
\sqrt{\frac{\xi(\omega)}{\left|\epsilon_p(\omega)\right|}}=
\begin{cases}
\cot\left(\frac{\kappa d}{2}\right) & (P=+),\\
-\tan\left(\frac{\kappa d}{2}\right) & (P=-).\label{kappa1}
\end{cases}
}
The Maxwell equations also lead to the additional constraint
\eqn{
\kappa=\sqrt{\xi(\omega)}\frac{\omega}{c}.\label{kappa2}
}
In the case of the frequency regime slightly below the bare phonon frequency $\Omega$ of paraelectric materials, the dielectric function $\xi(\omega)$ takes a singularly large value compared with $|\epsilon_p(\omega)|$. Thus, the conditions~\eqref{kappa1} and \eqref{kappa2}  lead to the approximate eigenmode equation
\eqn{
\sqrt{\xi(\omega)}\omega\simeq n\frac{\pi c}{d}\;\;\;\;\;(n=1,2,\ldots).
}
Owing to the singular nature of $\xi(\omega)$ in the limit of $\omega\to\Omega-0$, 
this condition can in principle be attained for an arbitrary integer $n$. It is this resonant structure that allows for generating a large number of soft-phonon modes in the heterostructure configuration.
In practice, when the damping of the phonon mode is included, there should exist an effective cutoff on possible values of $n$, but still many eigensolutions should be allowed as long as the damping rate is small enough in comparison with the phonon frequency.

\subsubsection{Effects of cavity losses}\label{loss_sec}

\begin{figure}[b]
\includegraphics[width=86mm]{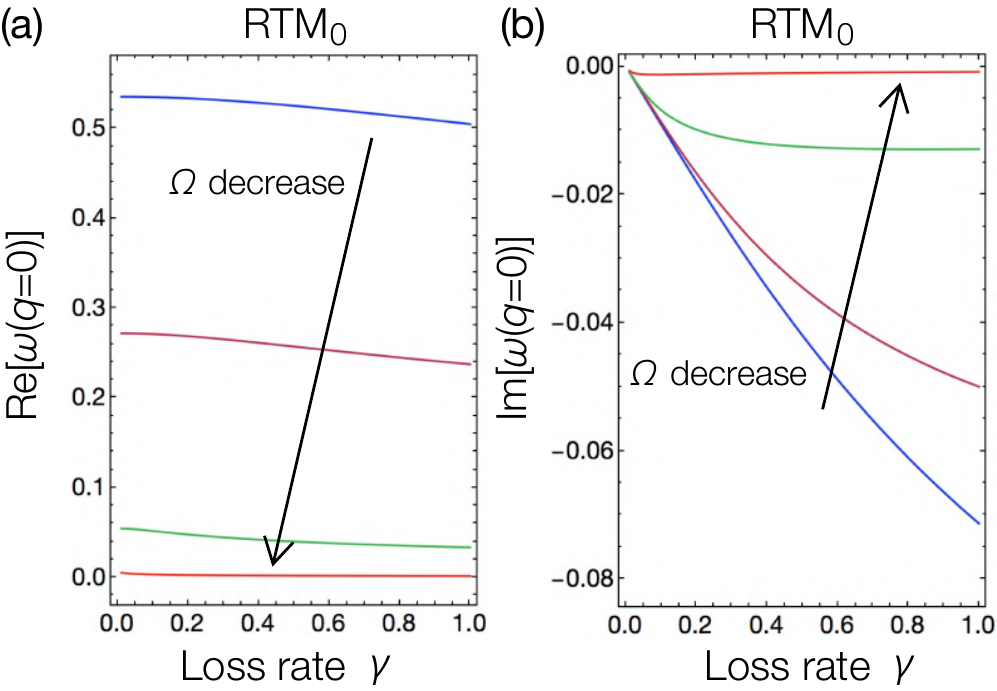} 
\caption{\label{fig_loss}
(a) Real and (b) imaginary parts of the complex eigenvalues for the most softened mode (i.e., RTM$_0$ mode) at $q=0$ in the presence of nonzero cavity-loss rate $\gamma$. From  top to bottom (resp. bottom to top) curves in (a)  (resp. (b)), the bare phonon frequency $\Omega$ is set to be  $1,\,0.5,\,0.1$, and $0.01$. The parameters are $d=1$, $\omega_p=5$, and $g=3.5$. 
}
\end{figure}

The analyses presented in the previous sections can be generalized to include additional complexities of experimental systems. In particular, in this section we  consider the effect of  scattering of electrons in metals by disorder, which can manifest itself as Ohmic losses. We use a phenomenological approach of adding an imaginary part to the permittivity of metallic mirrors. This corresponds to using the Drude model of electron dynamics, where damping effect is characterized by nonzero loss rate $\gamma$ (see Appendix~\ref{app_loss}). We use this model to  analyze the complex eigenvalues of the mode that exhibits the strongest softening, i.e., the RTM$_0$ mode. As shown in Fig.~\ref{fig_loss},  the mode frequency  deviates slightly from the corresponding value in the loss-less limit and now has a finite lifetime. However, when the bare phonon frequency approaches zero (i.e., $\Omega\to 0$), we find that not only the real part of the RTM$_0$ mode becomes zero, but also its imaginary part vanishes. This indicates that the lifetime of the softened mode diverges at the transition point. We emphasize that this conclusion holds true even for large $\gamma$. Thus, we conclude that the Ohmic losses in the cavity should not affect our main results, namely, the prediction of the cavity-induced softening and the resulting enhancement of ferroelectricity.

To understand the observed weak sensitivity to Ohmic losses in metals, we  recall that electric current in a metal is determined by the gradient of the electrochemical potential that includes both the electric field and the gradient of the chemical potential. A simple example of this is a problem of screening of a static charge inside a disordered metal; in this case, one finds finite electric field on the scale of a screening length, but no electric currents. This is also the reason why in our setup the system can develop static ferroelectric order without electric currents, i.e., no Ohmic losses. 

It is useful to provide another perspective on why losses do not affect the phase transition discussed here. In general, static impurities in metals, which cause Ohmic losses, should give rise to scattering between modes at the same energy, but different propagation directions and polarizations. If one launches a mode with a well defined momentum, scattering will lead to depletion in the population of this mode, which is understood as loss.
However, such scattering should not affect equilibrium thermodynamic properties of the system because it does not associate with actual dissipation of energy. More specifically, as the scattering modes have the same energy and thus equal populations in equilibrium, the detailed balance condition ensures equal rates for scattering processes going in the opposite directions, leading to no net dissipated energy. 

One may be concerned that this argument fails when inelastic scattering processes are included. However, as the RTM$_0$ mode frequency approaches zero, inelastic scattering processes become strongly suppressed because of the phase-space argument (e.g., the density of states of acoustic phonons scales as $\propto \omega^2$ at small frequencies). A more rigorous analysis of effects of dissipation can be performed using an effective model of coupling to a bosonic reservoir, or introducing the Green's function formalism, or the microscopic Hopfield approach. However, we expect that the conclusion of insensitivity of the phase transition to dissipation will not be affected.

\section{Cavity-enhanced ferroelectric phase transition in an interacting system\label{sec4}}
Building upon analysis of hybridized light-matter excitations in the cavity geometry, we now discuss the paraelectric-to-ferroelectric phase transition, which is the structural quantum phase transition commonly observed in ionic crystals. Our method of determining the phase transition point is based on identifying the softening of the dipolar phonon mode, which is a standard indication of the continuous second order phase transition into the broken symmetry phase. Interestingly, we find that in the cavity setting the transition point is shifted in favor of the ferroelectric phase. This softening results from the interplay between the modified dispersions and feedback effects through nonlinear interactions between the energetically dense elementary excitations.

\subsection{Variational principle}
Our primary objective in this section is to understand how the interaction term between the  phonon modes (cf. Eq.~\eqref{mimint}) modifies physical properties of quantum paraelectrics, especially in the vicinity of the transition into the ferroelectric phase. The paraelectric-to-ferroelectric phase transition is essentially a structural transition in an ionic crystal, where the phonon field $\hat{\boldsymbol \phi}$ plays the role of the order parameter. When it develops an expectation value, we find the ferroelectric phase characterized by spontaneous symmetry breaking. This corresponds to asymmetric distortion of charges within the unit cell, giving rise to a uniform electric polarization (cf. Fig.~\ref{fig1}(b)). In the bulk system, this transition is associated with  softening of the upper phonon-polariton mode, while the acoustic lower polariton mode is irrelevant as it reduces to a pure photon mode in the $q\rightarrow 0$ limit. In the cavity geometry of our interest, we thus need to analyze softening of the lowest optical mode, i.e., the RTM$_0$ mode.

We consider a system described by the Hamiltonian
\eqn{\label{HtotVar}
\hat{H}_{\rm tot}=\hat{H}_{\rm matter}+\hat{H}_{\rm light}+\hat{H}_{{\rm l}-{\rm m}},
}
where $\hat{H}_{\rm matter}$ now includes both the quadratic terms~\eqref{mimquad} and the most relevant interaction term~\eqref{mimint}, $\hat{H}_{\rm light}$ is the Hamiltonian for the free electromagnetic field~\eqref{emfree}, and $\hat{H}_{{\rm l}-{\rm m}}$ describes the light-matter coupling in the cavity setting as given by Eq.~\eqref{mimlm}. The inclusion of such nonlinear terms is particularly important close to the phase transition, where phonon fluctuations become significant. 

We use variational approach to analyze the phase diagram of the nonlinear model (\ref{HtotVar}). It is convenient to change notations to bring our model into the form commonly used in the study of second-order phase transitions. We replace the bare phonon frequency $\Omega^2$ in $\hat{H}_{0}$ (cf. Eq.~\eqref{mimquad}) with a parameter $r$, which corresponds to the tuning parameter of a physical system such as pressure, isotope concentration, or chemical composition. In the quadratic  (i.e., noninteracting) theory, paraelectric and ferroelectric phases correpond to $r\geq 0$ and $r<0$ respectively, while conditions for the interacting system will be discussed below.  
We separate the quadratic part of the total Hamiltonian as  $\hat{H}_{{\rm tot},0}(r)$. Then, the full interacting Hamiltonian can be written as  $\hat{H}_{{\rm tot}}(r)=\hat{H}_{{\rm tot},0}(r)+\hat{H}_{\rm int}$. The thermal equilibrium of the system is described by the usual Gibbs state
\eqn{\label{rhoGibbs}
\hat{\rho}(r)=\frac{e^{-\beta\hat{H}_{{\rm tot}}(r)}}{Z(r)}=\frac{e^{-\beta\left(\hat{H}_{{\rm tot},0}(r)+\hat{H}_{{\rm int}}\right)}}{Z(r)},
}
where $Z(r)\equiv {\rm Tr}[e^{-\beta\hat{H}_{{\rm tot}}(r)}]$. 
Within variational approach, we approximate the Hamiltonian in Eq.~(\ref{rhoGibbs}) by the quadratic Hamiltonian with an effective parameter  $r_{\rm eff}$ that should be determined from the variational principle: 
\eqn{\label{varrho}
\min\limits _{r_{\rm eff}}\;D_{{\rm KL}}\left[\hat{\rho}_{{0}}(r_{\rm eff})|\hat{\rho}(r)\right]\geq0,
}
where $D_{\rm KL}[\cdot|\cdot]$ is the Kullback-Leibler divergence \cite{ST13}, and an effective  Gaussian state is defined as
\eqn{
\hat{\rho}_{{0}}(r_{{\rm eff}})=\frac{e^{-\beta\hat{H}_{{\rm tot},0}(r_{{\rm eff}})}}{Z_{0}(r_{\rm eff})}
}
with $Z_{0}(r_{\rm eff})\equiv{\rm Tr}[e^{-\beta\hat{H}_{{\rm tot},0}(r_{{\rm eff}})}]$.

Our goal is to find $r_{\rm eff}$ that provides the best approximation to the interacting theory. To this end, we consider
\eqn{\label{varmain}
F_{v}&\equiv& F_0(r_{\rm eff})+\left\langle \hat{H}_{{\rm tot}}(r)-\hat{H}_{{\rm tot},0}(r_{\rm eff})\right\rangle _{{\rm eff}},
}
where we denote an expectation value with respect to the effective density matrix as $\langle\cdots\rangle_{{\rm eff}}={\rm Tr}[\hat{\rho}_{0}(r_{\rm eff})\cdots]$, and $F_{0}(r_{\rm eff})=-\frac{1}{\beta}\ln Z_{0}(r_{\rm eff})$ is the Gaussian variational free energy.  
The variational condition~\eqref{varrho} can then be simplified as \cite{RF72}
\eqn{
\min\limits _{r_{{\rm eff}}}F_{v}\geq F(r),
}
where $F(r)=-\frac{1}{\beta}\ln Z(r)$ is the exact free energy. The optimal effective tuning parameter $r_{\rm eff}^*$, which is defined by
\eqn{
r_{\rm eff}^* =\mathop{{\rm arg~min}}\limits _{r_{\rm eff}}F_v,
}
gives the best approximation of the interacting model in terms of the linear theory.
Within the present variational formalism, the phase transition point is determined from the condition $r_{\rm eff}^*=0$ that is achieved for a certain value of the bare tuning parameter $r$. When phonon nonlinearities are repulsive, the transition takes place for a negative value of the bare parameter $r$. In other words, the interaction feedback gives rise to the renormalization of the effective phonon frequency $r_{\rm eff}^*=\Omega_{\rm eff}^2$ in the variational theory $\hat{H}_{{\rm tot},0}(r_{\rm eff}^*)$, and the transition occurs when this renormalized frequency of paraelectric materials goes down to zero.

We note that the  variational method presented here is closely related to the standard mean-field approach, since we choose a simple Gaussian family of variational states. We approximate the free energy of an interacting  system by the free energy of a noninteracting system with an optimized variational parameter $r_{\rm eff}$. For this class of variational states, interaction terms in Eq.~(\ref{varmain}) can be calculated by factorizing them into the product of the expectation values of the quadratic operators with respect to the noninteracting theory, in the same manner as it is done in a standard mean-field theory  \cite{GNS75}  (see Appendix~\ref{app_vari} for technical details). We note that this theoretical approach is exact for the spherical model, in which the number of components in the order parameter is effectively infinite.

\begin{figure}[b]
\includegraphics[width=60mm]{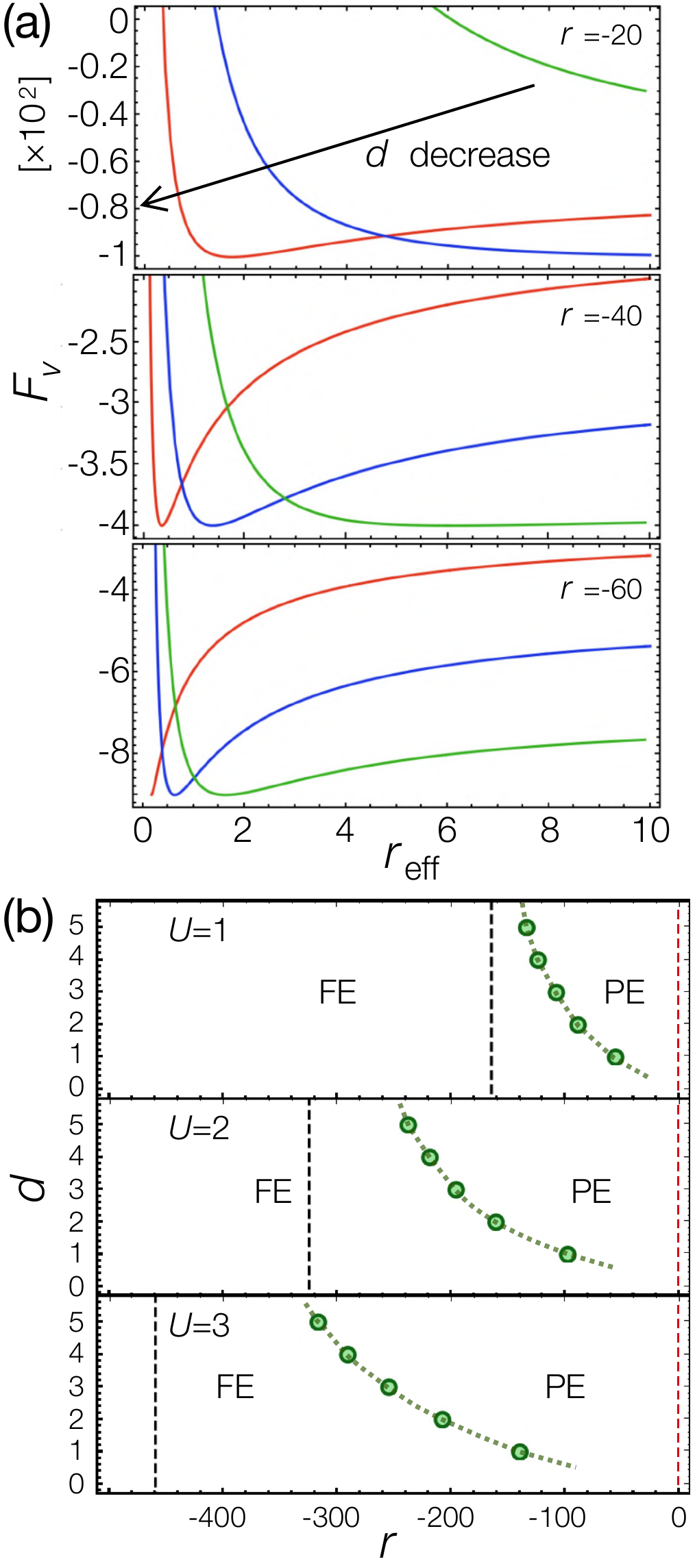} 
\caption{\label{fig9}
(a) Variational free energies $F_v$ plotted against the variational parameter $r_{\rm eff}$ with decreasing tuning parameter $r$ from top to bottom panels.  The red and blue solid curves correspond to the results for the cavity setting with the insulator thicknesses $d=1$ and $d=3$, respectively, while the green solid curves correspond to the bulk results. The minima of the variational energies determine the equilibrium values $r_{\rm eff}^*=\Omega_{\rm eff}^2$ of the effective phonon frequency including the nonlinear feedback. 
(b) Phase diagram in the cavity setting at different nonlinear strengths $U$. As the tuning parameter $r$ is decreased, the transition from the paraelectric phase (PE) to the ferroelectric phase (FE) occurs when the equilibrium effective phonon frequency $r_{\rm eff}^*$ goes down to zero. The green dashed curve is the phase boundary in the cavity geometry, while the black (resp. red) vertical dashed line is that in the bulk geometry (resp. at $U=0$).
We set $c=\hbar=\beta=1$, $\omega_p=5$, $g=3.5$, and the momentum cutoff to be $\Lambda_c=10^2$.
}
\end{figure}

\subsection{Results}

We show the obtained variational free energy $F_v$ against the variational parameter $r_{\rm eff}$ at different insulator thicknesses $d$ in Fig.~\ref{fig9}(a). From top to bottom panels, we decrease the bare tuning parameter $r$. The thermal equilibrium value $r_{\rm eff}^*$ of the effective phonon frequency  is determined from identifying the minimum of each variational free-energy landscape. In all the panels, a thinner insulator thickness $d$ in the cavity setting leads to a smaller equilibrium phonon frequency $r_{\rm eff}^*$, indicating the phonon softening owing to the coupling with light modes strongly confined in the cavity. 
With further decreasing the thickness $d$ or the tuning parameter $r$, the equilibrium phonon frequency eventually goes down to zero from above, $r_{\rm eff}^*\to0_+$, at which the paraelectric-to-ferroelectric phase transition occurs (see e.g., the red solid curve in the bottom panel of Fig.~\ref{fig9}(a)).

The corresponding phase diagram of the cavity setting with varying thickness $d$ is plotted in Fig.~\ref{fig9}(b) for different values of the phonon nonlinearity $U$. Here, the horizontal axis corresponds to the tuning parameter $r$, which may correspond to pressure or isotope concentration. In the paraelectric phase, decreasing the value of $r$ corresponds to approaching the transition point. We can then interpret the results in Fig.~\ref{fig9}(b) as the cavity-induced stabilization of the ferroelectric phase, which should be compared to the bulk transition point indicated by the black dashed line. We emphasize that this theoretical result has direct experimental implications. Consider, for example, a pressure-tuned transition into the ferroelectric phase in materials such as SrTiO$_3$ or KTaO$_3$, which are paraelectric at ambient conditions. Our results then predict that, when the sample is placed inside the cavity, transition into the ferroelectric phase should take place at lower pressure.

This favoring of the ferroelectric phase in the quantum electromagnetic environment is robust against changes in specific values of parameters and should be a generic feature of dipolar insulators confined in the cavities. We find that this effect becomes more pronounced for insulating films of smaller thickness $d$. We also observe that increasing the plasma frequency has qualitatively similar effect to decreasing the paraelectric slab thickness (cf. the inset in Fig.~\ref{fig3}(c)). This can be understood as a result of tighter mode confinement by metals with the higher plasma frequency.

The enhanced ferroelectricity found here can be understood from the interplay between the phonon nonlinearity and the dispersion of hybrid light-matter modes in the cavity geometry. To this end, we recall that the phonon repulsion included in our model renormalizes the effective phonon frequency as $\Omega_{\rm eff}=\sqrt{r_{\rm eff}^*}$. In general, low-energy modes provide the dominant contribution to the interaction-induced hardening of phonons. In the present cavity configuration, there are many low-lying dispersions, corresponding to different values of longitudinal wavenumber $\kappa$ for a fixed in-plane momentum $q$. The transition occurs when energy of the most-softened mode (RTM$_0$ mode) reaches zero, and it is sufficient that only this mode softens to have the ferroelectric transition. With decreasing the transverse size of the cavity $d$, the other modes (i.e., RTM$_{1,2,3,\ldots}$) are pushed to the frequency close to the frequency of the bare phonon, thus they contribute less to phonon hardening. In contrast, when we increase $d$, these modes go down in frequency and thus increase their contribution to phonon hardening, which is unfavorable to soften the RTM$_0$ mode, i.e., to induce the ferroelectric transition.
As shown in Fig.~\ref{fig9}(b), the enhanced ferroelectricity in the cavity configuration is more pronounced as the nonlinear strength $U$ is increased. In particular, we note that the enhancement is absent unless the nonlinearity is included into the analysis, as indicated by the red vertical line.

Theoretical predictions of our paper can be tested in several materials, including SrTiO$_3$ and KTaO$_3$ \cite{RSE14,EJ15}, as well as SnTe and Pb$_{1-z}$Sn$_z$Te \cite{WJA01}, where a tuning parameter can be varied by  pressure, chemical composition, or isotope substitution. This flexibility allows one to naturally prepare materials on the verge of the ferroelectric phase. Strong nonlinearity is also a ubiquitous feature of these materials. To be concrete, if we consider SrTiO$_3$, one can use the transverse optical phonon mode, for which the phonon nonlinear coupling is $U/M^2\!\simeq\!2.2\!\times \!10^3$\,$\rm THz$$^2$\,\AA$^{-2}$\,$\rm kg$$^{-1}$ \cite{KM19}. In the unit of $c=\hbar=\beta=1$ with, e.g., the temperature $T=60\,{\rm K}$, this nonlinear coupling strength becomes $U\simeq1.5$, which is close to the value used in the present numerical calculations \footnote{We note that the temperature is specified to be a concrete value in order to fix a length scale in the setup of interest.}. With this convention, the thickness $d=1$ corresponds to a length scale of several tens of microns. In this regime, the expected shift of the physical phonon frequency is an order of THz (cf. the bottom panel in Fig.~\ref{fig9}(a)), which should correspond to tens of degrees Kelvin shift in the transition temperature \cite{BD80,KM19}.

We remark that the model discussed here is still not sufficient to determine  spatial character of the ordered ferroelectric phase. From the symmetry of the softened RTM$_0$ mode, we expect that it will have an in-plane polarization, however, we cannot determine its momentum. It is common to address this question by analyzing unstable collective modes in the disordered (i.e., paraelectric) state and using $r$ that corresponds to the ordered (i.e., ferroelectric) phase. The wavevector of the most unstable modes is then an indication of the expected ordered state. In our consideration, this procedure  suggests instability at large $q$ (see also Appendix~\ref{app_hyb}). This is not surprising since dipoles have repulsive interactions when oriented side by side, thus dipolar interactions favor stripe domains with the modulation wavevector perpendicular to the direction of the dipoles. The tendency to having an ordered state at large  $q$ should be opposed by the elastic energy that favors the uniform value of $\phi$. This contribution can be described by the gradient term of $\phi$, which we did not include in our model. Only by properly including the competition between dipolar interactions and elastic energy, one can determine the correct ordering wavevector. We plan to discuss the nature of the ordered phase in a future publication.

\section{Summary and Discussions\label{sec5}}
Our work demonstrates a great promise of a heterostructure configuration as a platform towards controlling the quantum phase of matter by modifying  quantum electrodynamic environments in the absence of external pump. In particular, the present consideration suggests that the strong hybridization between the continuum of matter excitations and cavity electromagnetic fields enriches the spectrum of elementary excitations in a way qualitatively different from the bulk material. 
More specifically, the main results of this paper can be summarized as follows:
\begin{itemize}
\item[(i)]{{\it Analysis of hybridized collective modes in a cavity}. We analyzed the collective modes in the heterostructure configuration shown in Fig.~\ref{fig2}, i.e., the dielectric slab having an infrared
active phonon at frequency $\Omega$ surrounded by two metallic electrodes. Hybridization of the phonon modes of the dielectric, the cavity electromagnetic modes, and the plasmons of the surrounding metallic mirrors gives rise to a rich structure of the eigenmodes; in particular, we found the emergence of a number of energy modes at frequencies slightly below the bare phonon frequency $\Omega$. The origin of these eigenmodes can be understood from the strong frequency dependence of the dielectric function close to the phonon frequency (see Sec.~\ref{sec_resori}).
}
\item[(ii)]{{\it 
Cavity-enhanced ferroelectric phase transition}. To go beyond the simple quadratic model of the photon-phonon-plasmon hybridization, we included phonon nonlinearities in the dielectric material through the variational method. We found that the collective modes are softened as the thickness of the dielectric slab is decreased. We emphasize that this feature is present only when we include phonon nonlinearities and, therefore, its mechanism goes beyond simple screening of ferroelectric fluctuations by electrons in the metallic claddings.  The softening corresponds to the enhancement of the ferroelectric instability, and allows a cavity system to undergo transition into the ferroelectric phase even when the bulk material remains paraelectric. We suggest that such cavity-enhanced transition could be experimentally  observed in several quantum paraelectric materials at the verge of the ferroelectric phase.  
}
\end{itemize}

The mechanism proposed in this paper is not a specific feature of a particular setup, but should be applicable to a wide class of systems supporting excitations that possess electric dipole moments. For instance, our theory can  be used to analyze the coupling of cavity fields to excitonic molecules \cite{TJR05,ST11} and also to artificial quantum systems such as ultracold polar molecules \cite{SDI11}, trapped ions \cite{BV97}, and Rydberg atoms \cite{AGS84}. The present formulation should also be applicable to analyze polariton modes in the hexagonal boron nitride (hBN) \cite{DCR10,CG13,GA18}, which is a dielectric insulator material associated with a large optical phonon frequency. 
We point out again that the predicted changes of the elementary excitations  can occur without external driving in contrast to, e.g., the earlier studies of exciton-polaritons in microcavities \cite{YY02,KA07,JK06,Frank_2013,BT14,SD16,LA20}. 

As by-product of the analyses given in this paper, we envision several potential applications of the cavity-induced changes of the dispersive excitation modes.
For instance, one can consider a molecular crystal sandwiched between metallic mirrors, where  polaritons originate from the hybridization of singlet excitons and cavity electromagnetic modes. The corresponding triplet excitons have negligible interaction with light  and are thus unaffected by the cavity confinement. The heterostructure cavity configuration then allows one to change relative energies of singlet and triplet excitations, which can be applied to improve the efficiency of organic light-emitting devices and realize photon-exciton converters. We expect that the modification of the hybridized dispersion in the proposed cavity setup can also be used to control photochemical reactions in molecular systems. The similar physics may be observed in the presence of a single metal-dielectric interface. Further discussions can be found in Appendix~\ref{app4}.

Several open questions remain for future studies. 
First, we recall that the whole system considered in this paper is assumed to be in thermal equilibrium, i.e., the effective temperature is identical for both of the matter and the cavity. This assumption may not always apply; for example, the metal mirrors and the insulator could be coupled to external bath modes at different temperatures. Moreover, driving the cavity  into a high-intensity regime or coherent pumping of atoms should induce the cavity-mediated interactions among the underlying elementary excitations \cite{LID10,BK10,JB17,HJ17,NMA18}. It remains an intriguing question how our formalism can be generalized to such nonequilibrium situations. 

Second, a quantitative amount of the predicted modification of the elementary excitations and also that of the transition-point shift can in practice depend on several specifics such as finite thickness of metallic mirrors, the phonon-scattering rate in matter, and the plasma frequency and the conductivity in metals. It merits further study to explore the impact of those experimental complexities on the present results. 
In particular, it is intriguing to elucidate how dissipation effects on phonons and cavity photons can influence the phase diagram obtained in the present work. Our preliminary phenomenological analysis in Sec.~\ref{loss_sec} indicates that the equilibrium properties should not be qualitatively affected by dissipation; however, it may still cause quantitative modifications and, in particular, influence nonequilibrium dynamics. These issues could be  addressed by relying on recent developments in Markovian open quantum many-body systems \cite{DAJ14,Ashida20}. Even if this is not the case, one may in principle apply a more inclusive approach such as by explicitly including bath degrees of freedom as bosonic reservoirs into the analysis.  
Meanwhile, on the border of the phase transition, there are also interesting open questions on how quantum critical behavior is modified due to cavity confinement and how it can be affected by dissipative nature of a cavity  \cite{HJA76,*MAJ93,*TM85,MA06,PT08,DS10,YA17nc,*YA16,SLM13,RSE14,MJ16}. It would also be interesting to explore combining cavity-enhanced phase transitions with topological aspects of open systems \cite{CEB13,DH14,LL14,KT15,ZG18,BCE18}. 

Third, it will be intriguing to consider the situation in which the primary coupling between matter and light is mediated by the magnetic field. A concrete example can be Damon-Eschbach modes in ferromagnetic films \cite{RWD61}. The intrinsically chiral nature of these modes may qualitatively alter the character of collective modes in the cavity configuration. One can also consider a situation with both electric and magnetic couplings being important. This can be relevant to multiferroic systems \cite{EW06}, in which ferroelectric and magnetic orders are intertwined. 

Finally, it is also intriguing to consider further features and functionalities that can arise from fabricating additional structures at the metal-insulator interfaces. For instance, periodic texturing or distortion of the surfaces can lead to the formation of gaps in the in-plane dispersion as realized in photonic crystals. This will provide additional flexibility for manipulation of the hybridized light-matter excitations. 
It also merits further study to consider
how changes in the phonon spectrum in the metal-paraelectric heterostructure can affect many-body states in the metallic mirrors, such as, superconducting states. This should be particularly relevant to low-density superconductivity mediated by plasmons as discussed in the case of doped strontium titanate, where the soft transverse optical phonon plays a crucial role  \cite{RL16}. We hope that our work stimulates further studies in these directions.

\begin{acknowledgments}
We are grateful to Richard Averitt, Dmitri Basov, Antoine Georges, Bertrand I. Halperin, Mikhail Lukin, Dirk van der Marel, Giacomo Mazza, Marios Michael, Prineha Narang, Angel Rubio, and Sho Sugiura for fruitful discussions. Y.A. acknowledges support from the Japan Society for the Promotion of Science through Grant Nos.~JP16J03613 and JP19K23424, and Harvard University for hospitality. J.F. acknowledges funding from the Swiss National Science Foundation grant  200020-192330. 
D.J. acknowledges support by EPSRC grant No.~EP/P009565/1. D.J. and A.C. acknowledge funding from the European Research Council under the European Union's Seventh Framework Programme (FP7/2007-2013)/ERC Grant Agreement No.~319286 (Q-MAC). A.C. acknowledges support from the Deutsche Forschungsgemeinschaft (DFG) via the Cluster of Excellence `The Hamburg Centre for Ultrafast Imaging' (EXC 1074--Project ID 194651731) and from the priority program SFB925.
 E.D. acknowledges support from Harvard-MIT CUA, AFOSR-MURI Photonic Quantum Matter (award FA95501610323), and DARPA DRINQS program  (award D18AC00014).
\end{acknowledgments}

\appendix
\section{Diagonalization of the quadratic Hamiltonian}\label{app_diag}
Here we provide technical details about the diagonalization of the quadratic parts of the bulk Hamiltonian in the main text. To this end, we decompose the matter field as $\hat{\boldsymbol{\phi}}(\boldsymbol{r})=\hat{\boldsymbol{\phi}}^{\parallel}(\boldsymbol{r})+\hat{\boldsymbol{\phi}}^{\perp}(\boldsymbol{r})$ and $\hat{\boldsymbol{\pi}}(\boldsymbol{r})=\hat{\boldsymbol{\pi}}^{\parallel}(\boldsymbol{r})+\hat{\boldsymbol{\pi}}^{\perp}(\boldsymbol{r})$, whose explicit forms are given by  
\eqn{
\hat{\boldsymbol{\phi}}^\perp(\boldsymbol{r})&=&\sqrt{\frac{\hbar}{2MN\Omega}}\sum_{\boldsymbol{k}\lambda}\left(\hat{\phi}_{\boldsymbol{k}\lambda}^\perp e^{i\boldsymbol{k}\cdot\boldsymbol{r}}\boldsymbol{\epsilon}_{\boldsymbol{k}\lambda}+{\rm H.c.}\right),\label{phononcan}\\
\hat{\boldsymbol{\pi}}^\perp(\boldsymbol{r})&=&-i\sqrt{\frac{\hbar M\Omega}{2N}}\sum_{\boldsymbol{k}\lambda}\left(\hat{\phi}_{\boldsymbol{k}\lambda}^\perp e^{i\boldsymbol{k}\cdot\boldsymbol{r}}\boldsymbol{\epsilon}_{\boldsymbol{k}\lambda}-{\rm H.c.}\right),\label{phononmcan}\\
\hat{\boldsymbol{\phi}}^{\parallel}(\boldsymbol{r})&=&\sqrt{\frac{\hbar}{2MN\Omega}}\sum_{\boldsymbol{k}}\left(\hat{\phi}_{\boldsymbol{k}}^{\parallel}e^{i\boldsymbol{k}\cdot\boldsymbol{r}}\frac{\boldsymbol{k}}{|\boldsymbol{k}|}+{\rm H.c.}\right),\\
\hat{\boldsymbol{\pi}}^{\parallel}(\boldsymbol{r})&=&-i\sqrt{\frac{\hbar M\Omega}{2N}}\sum_{\boldsymbol{k}}\left(\hat{\phi}_{\boldsymbol{k}}^{\parallel}e^{i\boldsymbol{k}\cdot\boldsymbol{r}}\frac{\boldsymbol{k}}{|\boldsymbol{k}|}-{\rm H.c.}\right),
}
where $N=V/v$ is the number of modes,  $\hat{\phi}_{\boldsymbol{k}\lambda}^\perp$ ($\hat{\phi}_{\boldsymbol{k}\lambda}^{\perp\dagger}$) is the annihilation (creation) operator of transverse phonons with momentum $\boldsymbol k$ and polarization $\lambda$, and $\hat{\phi}_{\boldsymbol{k}}^\parallel$ ($\hat{\phi}_{\boldsymbol{k}}^{\parallel\dagger}$) is the annihilation (creation) operator of longitudinal phonons. We recall that the transverse polarization vectors ${\boldsymbol \epsilon}_{{\boldsymbol k}\lambda}$ satisfy Eqs.~\eqref{polarization} and \eqref{polarization2}. At the quadratic level ($U=0$), only the transverse components couple to the dynamical electromagnetic field while the longitudinal components couple to the static longitudinal electric field.  
In the similar manner as in the electromagnetic Hamiltonian $\hat{H}_{\rm light}$, the quadratic part $\hat{H}_0$ of the matter Hamiltonian $\hat{H}_{\rm matter}$ can then be diagonalized as in Eq.~\eqref{linmat} in the main text.

To diagonalize the transverse part of the total quadratic Hamiltonian $\hat{H}_{{\rm tot},0}^\perp$, it is useful to introduce the conjugate pairs of variables for the field operators of electromagnetic field $\hat{\boldsymbol \pi}^\perp$ and $\hat{\boldsymbol A}$ in the Fourier space:
 \eqn{
\hat{\pi}_{\boldsymbol{k}\lambda}^\perp&=&-i\sqrt{\frac{\hbar}{2\Omega}}\left(\hat{\phi}_{\boldsymbol{k}\lambda}^\perp-\hat{\phi}_{-\boldsymbol{k}\lambda}^{\perp\dagger}\right),\\
\hat{\pi}_{\boldsymbol{k}\lambda}^{\perp P}&=&\sqrt{\frac{\hbar\Omega}{2}}\left(\hat{\phi}_{\boldsymbol{k}\lambda}^\perp+\hat{\phi}_{-\boldsymbol{k}\lambda}^{\perp\dagger}\right),\\
\hat{A}_{\boldsymbol{k}\lambda}&=&\sqrt{\frac{\hbar}{2\omega_{\boldsymbol{k}}}}\left(\hat{a}_{\boldsymbol{k}\lambda}+\hat{a}_{-\boldsymbol{k}\lambda}^{\dagger}\right),\\
\hat{A}_{\boldsymbol{k}\lambda}^{P}&=&-i\sqrt{\frac{\hbar\omega_{\boldsymbol{k}}}{2}}\left(\hat{a}_{\boldsymbol{k}\lambda}-\hat{a}_{-\boldsymbol{k}\lambda}^{\dagger}\right),
 } 
 where the superscript $P$ indicates that an operator is a conjugate variable of the corresponding operator, i.e., each pair of conjugate variables satisfies the canonical commutation relations.  The transverse part of the quadratic Hamiltonian can then be simplified as 
 \eqn{
 \hat{H}_{{\rm tot},0}^\perp&\!=\!&\frac{1}{2}\sum_{\boldsymbol{k}\lambda}\!\left(\begin{array}{cc}
\hat{\pi}_{\boldsymbol{k}\lambda}^{\perp P} & \hat{A}_{\boldsymbol{k}\lambda}^{P}\end{array}\right)\left(\begin{array}{cc}
1 & 0\\
0 & 1
\end{array}\right)\left(\begin{array}{c}
\hat{\pi}_{-\boldsymbol{k}\lambda}^{\perp P}\\
\hat{A}_{-\boldsymbol{k}\lambda}^{P}
\end{array}\right)\nonumber\\
&\!+\!&\frac{1}{2}\sum_{\boldsymbol{k}\lambda}\!\left(\begin{array}{cc}
\hat{\pi}_{\boldsymbol{k}\lambda}^\perp & \hat{A}_{\boldsymbol{k}\lambda}\end{array}\right)\!\left(\begin{array}{cc}
\Omega^{2} & -g\Omega\\
-g\Omega & \omega_{\boldsymbol{k}}^{2}+g^{2}
\end{array}\right)\!\left(\begin{array}{c}
\hat{\pi}_{-\boldsymbol{k}\lambda}^\perp\\
\hat{A}_{-\boldsymbol{k}\lambda}
\end{array}\right),\nonumber\\
\label{quadH}
 }
from which one can readily obtain Eqs.~\eqref{totquadbulk} and \eqref{eigvbulk} in the main text.

\section{Macroscopic Maxwell equations}\label{app1}
In the linear regime, elementary excitations of the electromagnetic fields in nonabsorbing matter can in general be obtained by solving the macroscopic Maxwell equations with using boundary conditions appropriate for a physical setting of interest.  
Specifically, the dynamics of  the electromagnetic fields is governed by 
\eqn{
\nabla\times\boldsymbol{E}+\frac{\partial\boldsymbol{B}}{\partial t}&=&\boldsymbol{0},\\
\frac{1}{\mu_{0}}\nabla\times\boldsymbol{B}&=&\frac{\partial\boldsymbol{D}}{\partial t},\\
\nabla\cdot\boldsymbol{D}&=&0,\\
\nabla\cdot\boldsymbol{B}&=&0,
}
where $\boldsymbol{E}$ is the electric field, $\boldsymbol{B}$ is the magnetic field, $\mu_0$ is the vacuum permeability, and $\boldsymbol{D}$ is the macroscopic electric field in matter. Transforming to the frequency basis, the macroscopic electric field $\boldsymbol{D}$ is related to $\boldsymbol{E}$ via
\eqn{
\boldsymbol{D}\left(\boldsymbol{r}\right)=\epsilon_{0}\epsilon(\omega,\boldsymbol{r})\boldsymbol{E}\left(\boldsymbol{r}\right)
}
with $\epsilon_0$ being the vacuum permittivity and $\epsilon(\omega,\boldsymbol{r})$ being the frequency- and position-dependent relative permittivity reflecting material properties.  We choose the Coulomb gauge
 \eqn{
\nabla\cdot\boldsymbol{A}\left(\boldsymbol{r}\right)=0.
 }
In terms of the vector potential, the equation of motion can be simplified as 
 \eqn{\label{eigvA}
\frac{-\omega^{2}\epsilon\left(\omega,\boldsymbol{r}\right)}{c^{2}}\boldsymbol{A}\left(\boldsymbol{r}\right)+\nabla\times\left[\nabla\times\boldsymbol{A}\left(\boldsymbol{r}\right)\right]=\boldsymbol{0},
 }
 where $c=1/\sqrt{\epsilon_0\mu_0}$ is the vacuum light speed.
 In the bulk case, one can simply solve this eigenvalue problem by  employing the spatial invariance and expressing the eigensolutions in terms of the plane waves  (cf. Eq.~\eqref{planeA}). 
 
  {\renewcommand{\arraystretch}{1.5}
\begin{table*}
\caption{Eigenmodes in the cavity configuration. The labels of the bands are consistent with those shown in Fig.~\ref{fig6}. The polarization of each mode is specified by $\Lambda\in\{{\rm TM},{\rm TE},{\rm L}\}$, where TM and TE indicate the transverse magnetic and electric polarizations, respectively, and L denotes the longitudinal polarization. The parity of the mode functions ${\boldsymbol u}^\perp_\eta$ and ${\boldsymbol f}^\parallel_\eta$ are specified by $P=\pm$ (see also the main text). We also indicate whether or not the divergence of the electric field $\nabla\cdot\boldsymbol{{\cal E}}_{\eta}$ vanishes. The last column shows the explicit functional form of the mode function for each eigenmode, where ${\boldsymbol e}_{i}$ indicates the unit vector in the direction $i=x,y,z$.}\label{table2}
\begin{tabular}{@{\hspace{0mm}} l @{\hspace{5mm}} c @{\hspace{3.5mm}}c @{\hspace{3.5mm}} c @{\hspace{3.5mm}}l @{\hspace{0.5mm}}}
\hline
\hline
Bands&$\Lambda$ & $P$ & $\nabla\cdot\boldsymbol{{\cal E}}_{\eta}$ &\hspace{4.2cm}Mode function \\
\hline
\multirow{2}{*}{TM$_{2}$, RTM$_{n=0,2,4\ldots}$, SS ($=$TM$_{0}$)}&\multirow{2}{*}{TM} & \multirow{2}{*}{$+$} & \multirow{2}{*}{zero}  & $\boldsymbol{u}^\perp_{\eta}(z)=u_{\eta}^{{\rm I}}\left(\cos(\kappa_{\eta}z)\boldsymbol{e}_{x}-i\frac{q}{\kappa_{\eta}}\sin(\kappa_{\eta}z)\boldsymbol{e}_{z}\right)\left[\vartheta(d/2-z)+\vartheta(d/2+z)-1\right]$ \\
&&&&\hspace{1.2cm}$+u_{\eta}^{{\rm M}}\sum_{s=\pm1}e^{-s\nu_{\eta}z}\left(\boldsymbol{e}_{x}+is\frac{q}{\nu_{\eta}}\boldsymbol{e}_{z}\right)\vartheta(sz-d/2)$
 \\
 \multirow{2}{*}{TM$_{1}$, RTM$_{n=1,3,5\ldots}$, AS$_{1,2}$}&\multirow{2}{*}{TM} & \multirow{2}{*}{$-$} & \multirow{2}{*}{zero}  & $\boldsymbol{u}^\perp_{\eta}(z)=u_{\eta}^{{\rm I}}\left(\sin(\kappa_{\eta}z)\boldsymbol{e}_{x}+i\frac{q}{\kappa_{\eta}}\cos(\kappa_{\eta}z)\boldsymbol{e}_{z}\right)\left[\vartheta(d/2-z)+\vartheta(d/2+z)-1\right]$
 \\
&&&&\hspace{1.2cm}$+u_{\eta}^{{\rm M}}\sum_{s=\pm1}e^{-s\nu_{n}z}\left(s\boldsymbol{e}_{x}+i\frac{q}{\nu_{n}}\boldsymbol{e}_{z}\right)\vartheta(sz-d/2)$
 \\
 \multirow{2}{*}{TE$_{1}$, RTE$_{n=0,2,4\ldots}$}&\multirow{2}{*}{TE} & \multirow{2}{*}{$+$} & \multirow{2}{*}{zero}  & $\boldsymbol{u}^\perp_{\eta}(z)=u_{\eta}^{{\rm I}}\cos(\kappa_{\eta}z)\boldsymbol{e}_{y}\left[\vartheta(d/2-z)+\vartheta(d/2+z)-1\right]$
 \\
&&&&\hspace{1.2cm}$+u_{\eta}^{{\rm M}}\sum_{s=\pm1}e^{-s\nu_{\eta}z}\boldsymbol{e}_{y}\vartheta(sz-d/2)$
 \\
  \multirow{2}{*}{TE$_{0}$, RTE$_{n=1,3,5\ldots}$}&\multirow{2}{*}{TE} & \multirow{2}{*}{$-$} & \multirow{2}{*}{zero}  & $\boldsymbol{u}^\perp_{\eta}(z)=u_{\eta}^{{\rm I}}\sin(\kappa_{\eta}z)\boldsymbol{e}_{y}\left[\vartheta(d/2-z)+\vartheta(d/2+z)-1\right]$
 \\
&&&&\hspace{1.2cm}$+u_{\eta}^{{\rm M}}\sum_{s=\pm1}se^{-s\nu_{n}z}\boldsymbol{e}_{y}\vartheta(sz-d/2)$
 \\
 \multirow{2}{*}{L$_{\pm}$} & \multirow{2}{*}{L} & \multirow{2}{*}{$\pm$} & \multirow{2}{*}{nonzero}  & $\boldsymbol{f}^\parallel_{\eta}(z)=f_{\eta}^{{\rm I}}\left[\frac{q}{\sqrt{\kappa_\eta^2+q^2}}(e^{i\kappa_\eta z}\pm e^{-i\kappa_\eta z})\boldsymbol{e}_x+\frac{\kappa_\eta}{\sqrt{\kappa_\eta^2+q^2}}(e^{i\kappa_\eta z}\mp e^{-i\kappa_\eta z})\boldsymbol{e}_z\right]$
 \\
&&&&\hspace{1.2cm}$\times \left[\vartheta(d/2-z)+\vartheta(d/2+z)-1\right]$
 \\
\hline
\hline
\end{tabular}
\end{table*}

 \section{Eigenmodes in the cavity configuration}\label{app2}
 From now on, we focus on the heterostructure configuration as illustrated in Fig.~\ref{fig5}, which is symmetric under the spatial parity transformation along the direction perpendicular to the plane, i.e., under the reflection $(x,y,z)\to(x,y,-z)$. 
 \subsection{Transverse modes}
We first consider the transverse modes. In this case, the vector field can be quantized and expanded as
 \eqn{
 \hat{\boldsymbol{A}}(\boldsymbol{r})=\sum_{\eta}\sqrt{\frac{\hbar}{2\epsilon_{0}A\omega_{\eta}}}\left(\hat{a}_{\eta}\boldsymbol{{ U}}_{\eta}^\perp\left(\boldsymbol{r}\right)+{\rm H.c.}\right),
 }
 where $A$ is the area of the system, $\eta\in\left\{ \boldsymbol{q},n,\lambda\right\} $ labels each eigenmode in the cavity setting, $\hat{a}_\eta$ is its annihilation operator, $\omega_\eta$ is an eigenfrequency, and $\boldsymbol{{ U}}_{\eta}$ is the corresponding transverse eigenmode function satisfying $\nabla\cdot\boldsymbol{{ U}}_{\eta}^\perp=0$. Here, $\boldsymbol q$ is the two-dimensional in-plane wavevector, $n\in{\mathbb N}$ labels a discrete eigenmode in each sector $\lambda=(\Lambda,P)$ with $\Lambda\in\{{\rm TM},{\rm TE}\}$ specifying the transverse magnetic (TM) or the transverse electric (TE) polarization defined by the conditions,
\eqn{
\left[\nabla\times{\boldsymbol{{ U}}}_{\eta}^\perp({\boldsymbol r})\right]_z&=&{0}\;\;{\rm (TM\;modes)},\\
\left[{\boldsymbol{{ U}}}_{\eta}^\perp({\boldsymbol r})\right]_z&=&{0}\;\;{\rm (TE\;modes)},
}
and $P=\pm$ specifying the parity symmetry of the eigenmode function under the reflection,
\eqn{\label{parityAx}
[\boldsymbol{{ U}}_{\eta}^\perp(x,y,-z)]_{x,y}&=&P[\boldsymbol{{ U}}_{\eta}^\perp(x,y,z)]_{x,y},\\\label{parityAz}
[\boldsymbol{{ U}}_{\eta}^\perp(x,y,-z)]_z&=&-P[\boldsymbol{{ U}}_{\eta}^\perp(x,y,z)]_z.
}

Because of the spatial invariance in the $x$ and $y$ directions, we can decompose the eigenfunction as
 \eqn{
\boldsymbol{{ U}}_{\eta}^\perp\left(\boldsymbol{r}\right)=e^{i\boldsymbol{q}\cdot\boldsymbol{\rho}}\boldsymbol{u}^\perp_{\eta}(z),
 }
 where ${\boldsymbol \rho}=(x,y)^{\rm T}$.
The eigenvalue problem~\eqref{eigvA} in the present case can then be given by
\eqn{\label{eigvu}
\frac{-\omega_{\eta}^{2}\epsilon\left(\omega_{\eta},z\right)}{c^{2}}\boldsymbol{u}^\perp_{\eta}\left(z\right)\!+\!e^{-i\boldsymbol{q}\cdot\boldsymbol{\rho}}\nabla\times\left(\nabla\times\left[e^{i\boldsymbol{q}\cdot\boldsymbol{\rho}}\boldsymbol{u}^\perp_{\eta}\left(z\right)\right]\right)=\boldsymbol{0}\nonumber\\
}
with the relative permittivity being defined by
\eqn{
\epsilon(\omega,z)&=&\epsilon_{p}(\omega)\sum_{s=\pm}\vartheta(sz-d/2)\nonumber\\
&&+\xi(\omega)\left(\sum_{s=\pm}\vartheta(d/2-sz)-1\right).
}
Here, we use the Heaviside step function
\eqn{
\vartheta(x)=\begin{cases}
1 & \left(x>0\right)\\
0 & \left(x<0\right)
\end{cases},
}
and assume the relative permittivity functions in the metals ($|z|>d/2$) and the insulator ($|z|<d/2$) as
\eqn{\label{metdiele}
\epsilon_{p}(\omega)&=&1-\frac{\omega_{p}^{2}}{\omega^{2}},\\
\xi(\omega)&=&1+\frac{g^{2}}{\Omega^{2}-\omega^{2}}.\label{insdiele}
}
We recall that $\omega_p$ is the plasma frequency in the metal mirrors, $\Omega$ is the frequency of the nondispersive matter excitation such as an optical phonon mode in ionic crystals, and $g$ characterizes the strength of the light-matter coupling between the  matter dipolar mode and cavity electromagnetic fields. 
It is also useful to express the electric and magnetic fields in terms of the eigenmodes as follows:
\eqn{
\hat{\boldsymbol{E}}(\boldsymbol{r})&=&\sum_{\eta}\sqrt{\frac{\hbar\omega_{\eta}}{2\epsilon_{0}A}}\left(\hat{a}_{\eta}\boldsymbol{{\cal E}}_{\eta}\left(\boldsymbol{r}\right)+{\rm H.c.}\right),\\
\boldsymbol{{\cal E}}_{\eta}\left(\boldsymbol{r}\right)&=&i\boldsymbol{{ U}}^\perp_{\eta}\left(\boldsymbol{r}\right),\\
\hat{\boldsymbol{B}}(\boldsymbol{r})&=&\sum_{\eta}\sqrt{\frac{\hbar}{2\epsilon_{0}A\omega_{\eta}}}\left(\hat{a}_{\eta}\boldsymbol{{\cal B}}_{\eta}\left(\boldsymbol{r}\right)+{\rm H.c.}\right),\\
\boldsymbol{{\cal B}}_{\eta}\left(\boldsymbol{r}\right)&=&\nabla\times\boldsymbol{{ U}}^\perp_{\eta}\left(\boldsymbol{r}\right).
}
The boundary conditions require the continuity of ${\boldsymbol n}\times \boldsymbol{{\cal E}}_{\eta}$ and ${\boldsymbol n}\cdot[\epsilon_0\epsilon \boldsymbol{{\cal E}}_{\eta}]$ across the interfaces with $\boldsymbol n$ being the unit vector perpendicular to the surfaces. The latter condition is equivalent to the continuity of the magnetic field orthogonal to the surface normal ${\boldsymbol n}\times \boldsymbol{{\cal B}}_{\eta}$.

\subsubsection{TM modes}
 Without loss of generality,  we here consider the TM-polarized eigenmodes satisfying
 \eqn{
\left[\boldsymbol{{\cal B}}_{\eta}\right]_y&\neq&0,\;\;\left[\boldsymbol{{\cal B}}_{\eta}\right]_{x,z}=0,\\
\left[\boldsymbol{{\cal E}}_{\eta}\right]_y&=&0,\;\;\left[\boldsymbol{{\cal E}}_{\eta}\right]_{x,z}\neq0.
 } 
 The Maxwell equation $\nabla\cdot\hat{\boldsymbol B}=0$ then leads to $q_y=0$, i.e., the eigenmodes propagate along the $x$ axis and thus we denote $q=q_x$. The parity conditions~\eqref{parityAx} and \eqref{parityAz} in the present case can be read as
 \eqn{
 \left[\boldsymbol{{\cal B}}_{\eta}(x,y,-z)\right]_{y}&=&-P\left[\boldsymbol{{\cal B}}_{\eta}(x,y,z)\right]_{y},\\
\left[\boldsymbol{{\cal E}}_{\eta}(x,y,-z)\right]_{x}&=&P\left[\boldsymbol{{\cal E}}_{\eta}(x,y,z)\right]_{x},\\
\left[\boldsymbol{{\cal E}}_{\eta}(x,y,-z)\right]_{z}&=&-P\left[\boldsymbol{{\cal E}}_{\eta}(x,y,z)\right]_{z}.
 }
 In terms of the mode function ${\boldsymbol u}^\perp_\eta$, the TM modes considered here satisfy  
 \eqn{
\left[\boldsymbol{u}^\perp_{\eta}(z)\right]_{y}&=&0,\\
\left[\boldsymbol{{u}}^\perp_{\eta}(-z)\right]_{x}&=&P\left[\boldsymbol{{u}}^\perp_{\eta}(z)\right]_{x},\\
\left[\boldsymbol{{u}}^\perp_{\eta}(-z)\right]_{z}&=&-P\left[\boldsymbol{{u}}^\perp_{\eta}(z)\right]_{z}.
 }

\begin{figure*}[t]
\includegraphics[width=172mm]{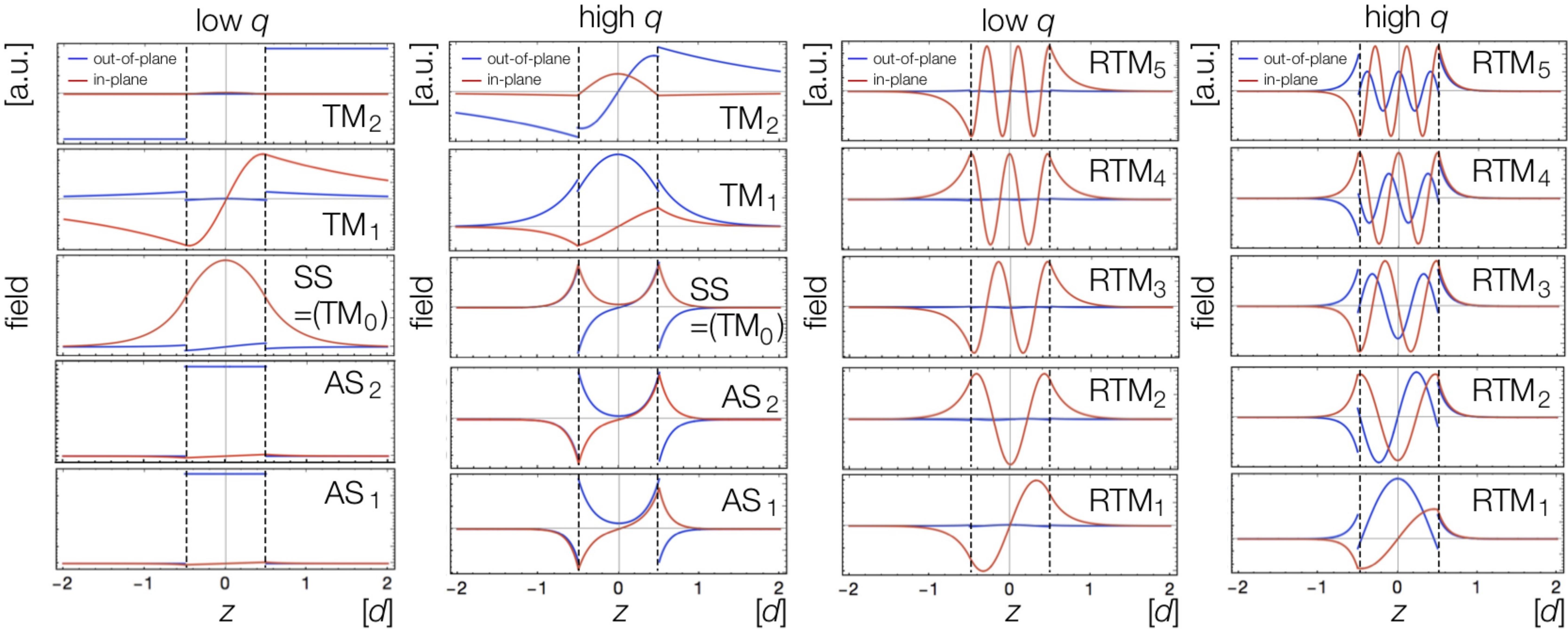} 
\caption{\label{figapp1}
Spatial profiles of the TM modes in the cavity configuration at low and high in-plane momenta $q$. The black dashed vertical lines indicate the metal-insulator interfaces. The blue solid curve shows the spatial profile of the out-of-plane component $[i{\boldsymbol u}^\perp_{\eta}(z)]_{z}$ of the mode function (corresponding to the $z$ component of the electric field). The red solid curve shows the spatial profile of the in-plane component $[{\boldsymbol u}^\perp_{\eta}(z)]_{x}$ of the mode function (corresponding to the $x$ component of the electric field). 
The parameters are the same as in Fig.~\ref{fig6} while the in-plane momentum $q$ for low- and high-momentum results is chosen to be $q=0.1\,\Omega/c$ and $q=7\,\Omega/c$, respectively.
}
\end{figure*}
 
 We first consider the case of the even parity $P=+$.
 Solving the eigenvalue equation~\eqref{eigvu} for the mode function ${\boldsymbol u}^\perp_\eta$ with the transversality condition, i.e., $\nabla\cdot (e^{i\boldsymbol{q}\cdot\boldsymbol{\rho}}\boldsymbol{u}^\perp_{\eta})=0$, we obtain the explicit functional form of ${\boldsymbol u}^\perp_\eta$ for this class of eigenmodes as shown in the first line  of Table~\ref{table2}. This set includes the bands labeled as TM$_{2}$, RTM$_{n=0,2,4\ldots}$, SS ($=$TM$_{0}$) that are shown in Fig.~\ref{fig6}(a,d). In these modes, the longitudinal wavenumber $\kappa_\eta$ in the insulator region is defined by 
\eqn{
\kappa_{\eta}^{2}=\frac{\omega_{\eta}^{2}}{c^{2}}\xi\left(\omega_{\eta}\right)-q^{2}.
} 
We emphasize that $\kappa_\eta$ takes either a real value or a purely imaginary value. In the latter case, the eigenmodes are localized at the interfaces and represent the surface modes. 
We define the skin wavenumber $\nu_\eta$, which is  real and positive,  as follows:
 \eqn{
\nu_{\eta}=\sqrt{q^{2}-\frac{\omega_{\eta}^{2}}{c^{2}}\epsilon_{p}\left(\omega_{\eta}\right)}>0.
 }
 The coefficients $u_\eta^{\rm M}$ and $u_\eta^{\rm I}$ in the metals (M) and in the insulator (I), respectively, satisfy the following relations imposed by the boundary conditions:
 \eqn{
u_{\eta}^{{\rm M}}/u_{\eta}^{{\rm I}}&=&e^{\nu_{\eta}d/2}\cos\left(\kappa_{\eta}d/2\right)\nonumber\\
&=&-\frac{\xi\left(\omega_{\eta}\right)\nu_{\eta}}{\epsilon_{p}\left(\omega_{\eta}\right)\kappa_{\eta}}e^{\nu_{\eta}d/2}\sin\left(\kappa_{\eta}d/2\right).
 }
  Aside an irrelevant phase factor, these coefficients can be fixed by further imposing the normalization condition $\int dz\,\left|\boldsymbol{u}^\perp_{\eta}\left(z\right)\right|^{2}=1$.

In the same way as done above, we can also obtain the corresponding eigenmodes in the case of the odd parity $P=-$.  Their functional forms are given in the second line in Table~\ref{table2}. The corresponding coefficients satisfy the boundary conditions
\eqn{
u_{\eta}^{{\rm M}}/u_{\eta}^{{\rm I}}&=&e^{\nu_{\eta}d/2}\sin\left(\kappa_{\eta}d/2\right)\nonumber\\
&=&\frac{\xi\left(\omega_{\eta}\right)\nu_{\eta}}{\epsilon_{p}\left(\omega_{\eta}\right)\kappa_{\eta}}e^{\nu_{\eta}d/2}\cos\left(\kappa_{\eta}d/2\right),
}
and also the normalization condition $\int dz\,\left|\boldsymbol{u}^\perp_{\eta}\left(z\right)\right|^{2}=1$.
These modes are labeled as TM$_{1}$, RTM$_{n=1,3,5\ldots}$, AS$_{1,2}$ that are shown in Fig.~\ref{fig6}(b,e).  
We show typical spatial profiles of the electric fields at low and high in-plane momenta $q$ for each of these TM modes in Fig.~\ref{figapp1}.  

\subsubsection{TE modes}

In the case of the TE polarization, without loss of generality we can assume the following conditions:
 \eqn{
\left[\boldsymbol{{\cal E}}_{\eta}\right]_y&\neq&0,\;\;\left[\boldsymbol{{\cal E}}_{\eta}\right]_{x,z}=0,\\
\left[\boldsymbol{{\cal B}}_{\eta}\right]_y&=&0,\;\;\left[\boldsymbol{{\cal B}}_{\eta}\right]_{x,z}\neq0.
 } 
From the transversality condition, 
\eqn{\nabla\cdot (e^{i\boldsymbol{q}\cdot\boldsymbol{\rho}}\boldsymbol{u}^\perp_{\eta})=0,} 
and the Maxwell equation $\nabla\cdot\hat{\boldsymbol D}=0$, we obtain the condition $q_y=0$, i.e., the TE modes considered here also propagate along the $x$ axis. The electromagnetic fields of the TE modes transform under the parity transformation as
 \eqn{
 \left[\boldsymbol{{\cal E}}_{\eta}(x,y,-z)\right]_{y}&=&P\left[\boldsymbol{{\cal E}}_{\eta}(x,y,z)\right]_{y},\\
 \left[\boldsymbol{{\cal B}}_{\eta}(x,y,-z)\right]_{x}&=&-P\left[\boldsymbol{{\cal B}}_{\eta}(x,y,z)\right]_{x},\\
\left[\boldsymbol{{\cal B}}_{\eta}(x,y,-z)\right]_{z}&=&P\left[\boldsymbol{{\cal B}}_{\eta}(x,y,z)\right]_{z}.
 }
 These conditions can be read in terms of the mode function ${\boldsymbol u}^\perp_\eta$ as 
 \eqn{
\left[\boldsymbol{u}^\perp_{\eta}(z)\right]_{x,z}&=&0,\\
\left[\boldsymbol{{u}}^\perp_{\eta}(-z)\right]_{y}&=&P\left[\boldsymbol{{u}}^\perp_{\eta}(z)\right]_{y}.
 }
 The eigenvalue equation then leads to the functional form in the third (fourth) line in Table~\ref{table2} in the case of the even parity $P=+$ (the odd parity $P=-$).  
The boundary conditions require that the value of the mode function ${\boldsymbol u}^\perp_\eta$ and its first spatial derivative must be continuous across the interfaces. More specifically, in the case of the even parity they lead to the conditions on the coefficients $u_{\eta}^{\rm M,I}$, 
\eqn{
u_{\eta}^{{\rm M}}/u_{\eta}^{{\rm I}}&=&e^{\nu_{\eta}d/2}\cos\left(\kappa_{\eta}d/2\right)\nonumber\\
&=&-\frac{\nu_{\eta}}{\kappa_{\eta}}e^{\nu_{\eta}d/2}\sin\left(\kappa_{\eta}d/2\right),
} 
while in the odd-parity case we have to impose the conditions
\eqn{
u_{\eta}^{{\rm M}}/u_{\eta}^{{\rm I}}&=&e^{\nu_{\eta}d/2}\sin\left(\kappa_{\eta}d/2\right)\nonumber\\
&=&\frac{\nu_{\eta}}{\kappa_{\eta}}e^{\nu_{\eta}d/2}\cos\left(\kappa_{\eta}d/2\right).
}
The normalization condition $\int dz\,\left|\boldsymbol{u}^\perp_{\eta}\left(z\right)\right|^{2}=1$ can be also imposed on the coefficients. In Fig.~\ref{fig6}(c,f), the TE modes with the even parity are labeled as TE$_{1}$, RTE$_{n=0,2,4\ldots}$ while the ones with the odd parity are labeled as  TE$_{0}$, RTE$_{n=1,3,5\ldots}$.
In contrast to some of the TM modes (such as the AS$_{1,2}$ and the RTM$_0$ modes), the spatial profiles of the electric fields are qualitatively insensitive to the in-plane momenta for all of the TE modes; their typical results are shown in Fig.~\ref{figapp2}.

\subsection{Longitudinal mode}
The other type of solutions is the longitudinal mode that is characterized by the longitudinal electric field  
\eqn{
\nabla\times\boldsymbol{{\cal E}}_\eta= 0.} 
Because of our choice of the Coulomb gauge $\hat{\boldsymbol A}^\parallel={\boldsymbol 0}$, the longitudinal field is directly linked with the longitudinal mode of the matter field $\hat{\boldsymbol \phi}^\parallel$ as discussed in the main text (cf. Eq.~\eqref{gaugeconst2}). From the Maxwell equation $\nabla\cdot\hat{{\boldsymbol D}}=0$, we can see that this mode oscillates in time at a constant frequency $\Omega^\parallel$ corresponding to the vanishing permittivity in the insulator
\eqn{
\xi\left(\Omega^{\parallel}\right)&=&0,\\
\Omega^{\parallel}&=&\sqrt{\Omega^{2}+g^{2}}.
}
We can expand the longitudinal matter field in terms of the mode function as
\eqn{
\hat{\boldsymbol{\phi}}^{\parallel}(\boldsymbol{r})=\sqrt{\frac{\hbar}{2MN_{d}\Omega}}\sum_{\eta}\left(\hat{\phi}_{\boldsymbol{q}n\lambda}^{\parallel}e^{i\boldsymbol{q}\cdot\boldsymbol{\rho}}\boldsymbol{f}^\parallel_{\eta}(z)+{\rm H.c.}\right),\nonumber \\
}
where  the longitudinal mode function satisfies  
\eqn{
\nabla\times(e^{i\boldsymbol{q}\cdot\boldsymbol{\rho}}\boldsymbol{f}^\parallel_{\eta}(z))=\boldsymbol{0}.}  
Meanwhile, the Maxwell equation $\nabla\times\hat{\boldsymbol E}=-\frac{\partial\hat{\boldsymbol B}}{{\partial t}}$ ensures the vanishing magnetic field $\hat{\boldsymbol B}={\boldsymbol 0}$. Thus, it suffices to only consider the longitudinal electric field and, without loss of generality, we here consider the modes satisfying
 \eqn{
\left[\boldsymbol{{\cal E}}_{\eta}\right]_y&=&0,\;\;\left[\boldsymbol{{\cal E}}_{\eta}\right]_{x,z}\neq0,\\
\boldsymbol{{\cal B}}_{\eta}&=&{\boldsymbol 0}.
} 
The corresponding in-plane momenta point to the $x$ direction, i.e., $q_y=0$ and thus $q=q_x$. 
The boundary conditions at the interfaces lead to the vanishing electromagnetic fields in the metal mirrors. The resulting explicit functional form of the mode function is given in the last line in Table~\ref{table2}. Depending on the parity $P=\pm$ defined via the relations
 \eqn{
\left[\boldsymbol{f}^\parallel_{\eta}(z)\right]_{y}&=&0,\\
\left[\boldsymbol{f}^\parallel_{\eta}(-z)\right]_{x}&=&P\left[\boldsymbol{f}^\parallel_{\eta}(z)\right]_{x},\\
\left[\boldsymbol{f}^\parallel_{\eta}(-z)\right]_{z}&=&-P\left[\boldsymbol{f}^\parallel_{\eta}(z)\right]_{z},
 }
 there are two classes of solutions, which are labeled as the L$_\pm$ modes in Fig.~\ref{fig6}(a). The corresponding longitudinal wavenumber $\kappa_\eta$ can be simply obtained by the boundary conditions $\left[\boldsymbol{{f}}^\parallel_{\eta}(\pm d/2)\right]_{x}=0$.

\begin{figure}[t]
\includegraphics[width=86mm]{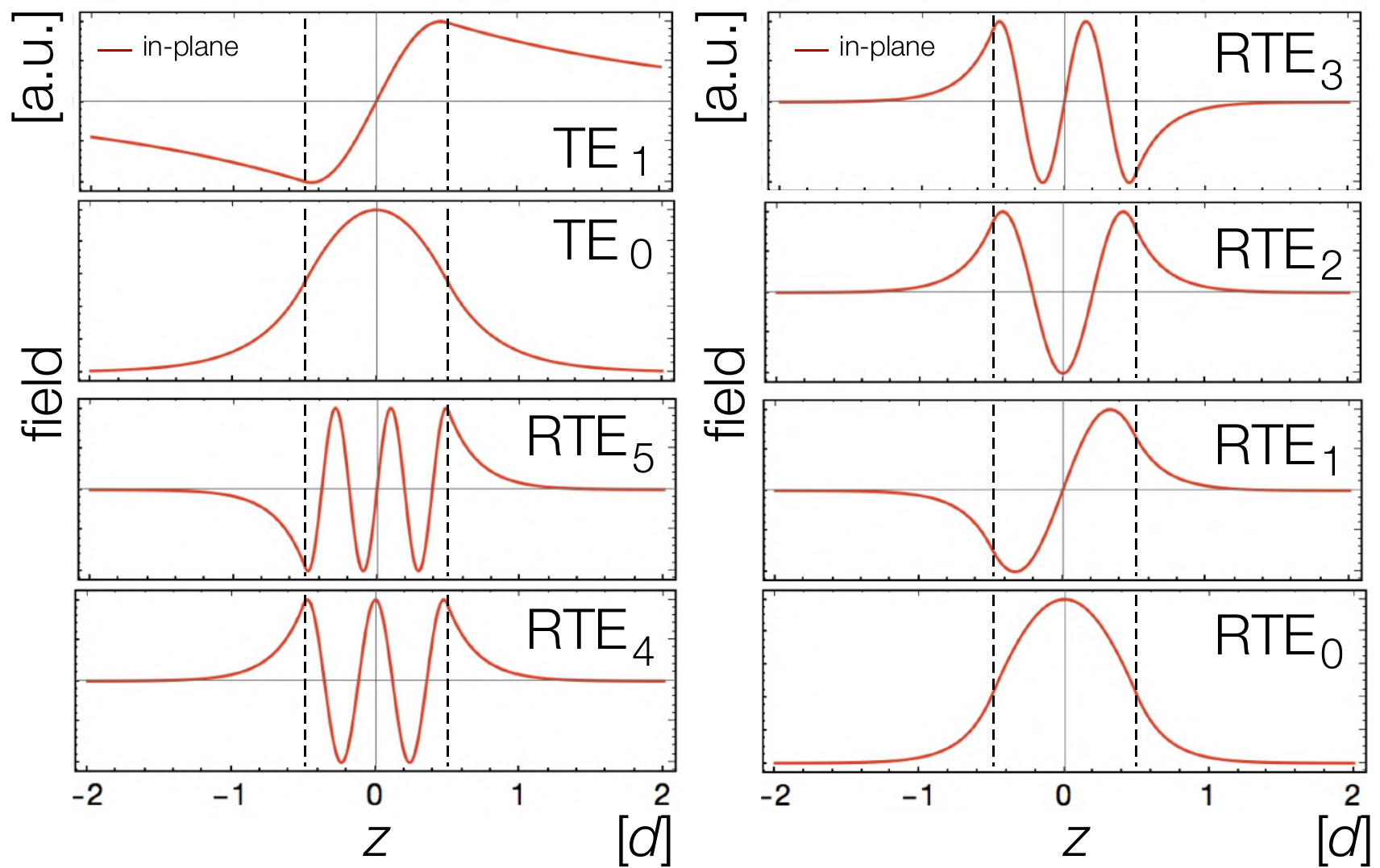} 
\caption{\label{figapp2}
Spatial profiles of the TE modes in the cavity configuration. The black dashed vertical lines indicate the metal-insulator interfaces.
The mode functions ${\boldsymbol f}^\parallel_\eta$ of all the TE modes have only the in-plane components, whose spatial profiles are shown by the red solid curve  (corresponding to the $y$ component of the electric field). 
The parameters are the same as in Fig.~\ref{fig6}. 
We use $q=0.1\,\Omega/c$ for the sake of concreteness while the spatial profiles of the TE modes are qualitatively insensitive to a specific choice of the in-plane momentum $q$.
}
\end{figure}

\section{Loss effect on the softened polariton mode\label{app_loss}}
We here provide further details about the analysis of the cavity loss. To be specific, we employ the Drude model and use a phenomenological complex permittivity. The resulting complex eigenvalue equation at $q=0$ is
\eqn{\label{eigvloss}
\omega=\frac{\kappa\sqrt{\omega_{p}^{2}/(\omega^{2}+i\omega\gamma)-1}}{\xi(\omega)\tan(\kappa d/2)},
}
where we recall that $\gamma>0$ represents the cavity-loss rate. The results in Fig.~\ref{fig_loss} in the main text are obtained by solving Eq.~\eqref{eigvloss} for the solution that reduces to the RTM$_0$ mode in the loss-less limit $\gamma\to 0$. 

Two remarks are in order. Firstly, while the imaginary part of the complex Drude permittivity diverges in $\omega\to 0$, this does not mean that an eigenmode should have a vanishingly short lifetime. The reason is that an eigenfrequency can go down to zero if and only if the bare phonon frequency also vanishes $\Omega\to 0$, indicating that the divergence of the Drude permittivity is necessarily compensated by that of $\xi(\omega)$ (cf. Eq.~\eqref{eigvloss}). Secondly, it is worthwhile to note that the Kramers-Kronig relations can in principle be recovered even in the loss-less limit by extending the permittivity as follows \cite{AZ65}:
\eqn{
\epsilon_{p}(\omega)=1-\omega_{p}^{2}\left(\frac{1}{\omega^{2}}+i\pi\delta^{(1)}(\omega)\right),
}
where $\delta^{(1)}$ is the first derivative of the Dirac delta function. While this additional contribution can be of physical importance when one is interested in real-time dynamics, it is irrelevant to the main results discussed in the present paper.

\section{Details about physical properties of the hybridized modes  \label{app_hyb}}

\begin{figure}[t]
\includegraphics[width=86mm]{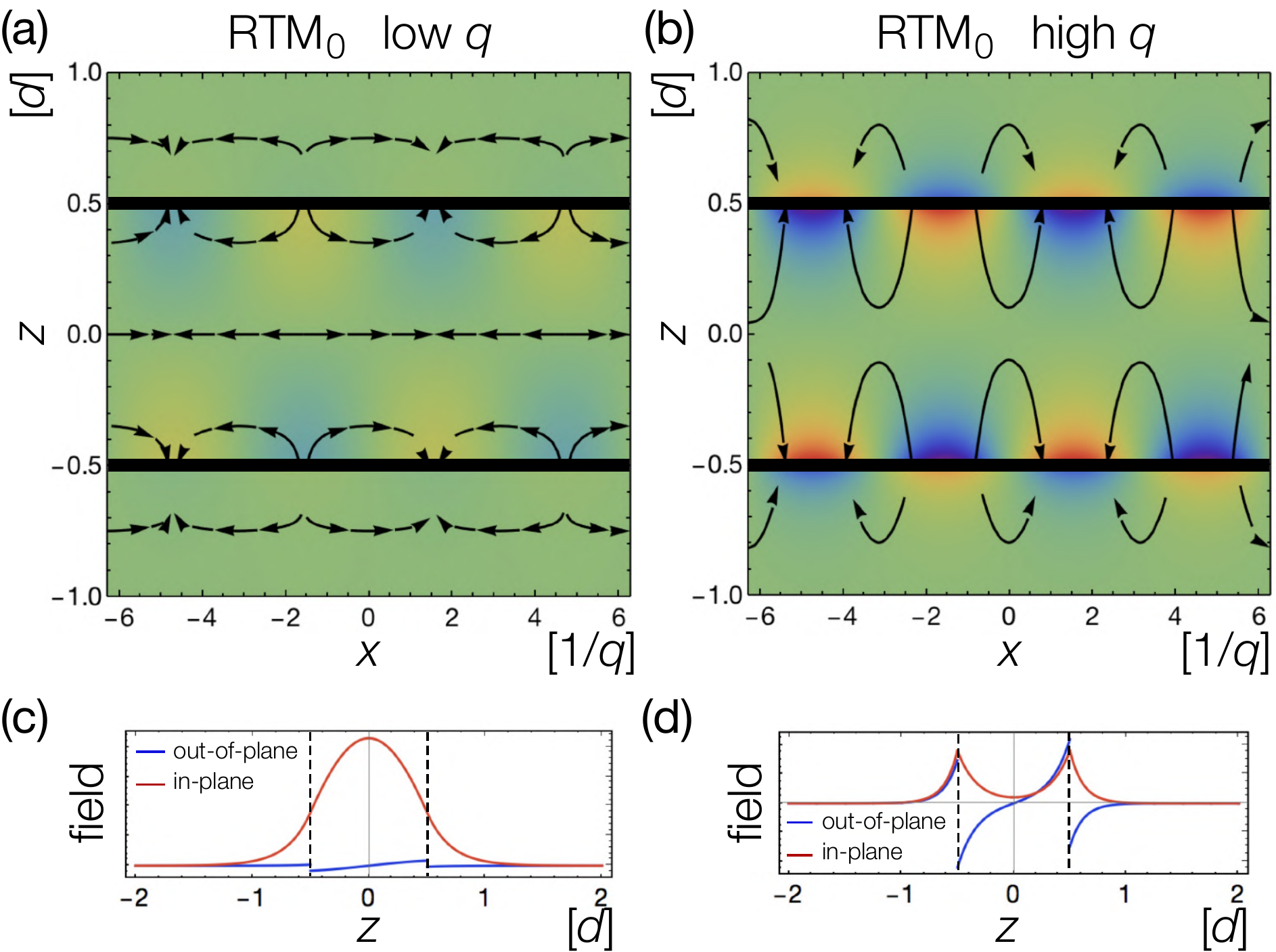} 
\caption{\label{fig7}
(a,b) Electrical flux lines and (c,d) spatial profiles of the mode function (i.e., the electric field) for the most-softened phonon mode denoted as the RTM$_0$ mode. Without loss of generality, we consider the mode propagating in the $x$ direction with the in-plane momentum $q=q_x$ and the vanishing electric field in the $y$ direction $E_y=0$. 
(a,b) The black solid curved arrows indicate the electrical flux lines on the $xz$ plane at (a) low and (b) high in-plane momenta $q$. The color indicates the magnitude of the electric field $E_z$ perpendicular to the interfaces.  (c,d) The spatial profiles of the in-plane (red solid curve) and the out-of-plane (blue solid curve) components of the electric field at $x=0$ in (a) and (b) are plotted in (c) and (d) in the arbitrary unit, respectively. The black dashed vertical lines indicate the interfaces at $|z|=d/2$. 
The parameters are $d=c/\Omega$, $g=3.5\,\Omega$, and $\omega_p=5\,\Omega$. We set $q=0.1\,\Omega/c$ in (a,c) and $q=7\,\Omega/c$ in (b,d). 
}
\end{figure}
We here discuss full details about physical properties of the hybridized elementary modes. Firstly, to gain further insights into the most-softened mode, in Fig.~\ref{fig7} we plot the spatial profiles of the electric field $\boldsymbol E$ of the RTM$_0$ mode. Figures~\ref{fig7}(a) and (b) show the electrical flux lines of the RTM$_0$ mode at low and high in-plane momenta $q$, respectively, with the color indicating the magnitude of the out-of-plane component $E_z$ of the electric field. We note that our convention of denoting the $z$-component as the out-of-plane one is in accordance with the planar cavity setting whose metal-insulator interfaces are parallel to the $xy$ plane. Without loss of generality, here we focus on the mode propagating in the $x$ direction, i.e. $q=q_x$, leading to the condition $E_y=0$ (see Appendix~\ref{app2}). Figures~\ref{fig7}(c) and (d) show the low-$q$ and high-$q$ spatial profiles of the in-plane (red solid curve) and the out-of-plane (blue solid curve) components of the electric field at $x=0$. At low $q$ before crossing the lower bulk  dispersion, the field predominantly points to the in-plane direction and extends over the insulator region $|z|\leq d/2$ (Fig.~\ref{fig7}(c)). In contrast, after crossing the bulk dispersion, the longitudinal wavenumber $\kappa_\eta$ turns out to take a purely imaginary value and the field profiles qualitatively change. Specifically, at high $q$ the fields are localized around the interfaces and the $z$ component of the electric field exhibits significant discontinuities due to the surface charges concentrated on the metal interfaces (Fig.~\ref{fig7}(d)); these features are characteristic of surface polariton modes. As shown in the main text, it is the softening of the RTM$_0$ mode discussed here that will trigger the structural instability of materials, thus ultimately inducing the paraelectric-to-ferroelectric phase transition.

We next consider the symmetric and antisymmetric surface modes denoted as SS(=TM$_0$) and AS$_2$ in Figs.~\ref{fig6}(a) and (b), respectively. 
The assignment of the label TM$_0$ to the SS mode is in accordance with the convention of studies in plasmonics \cite{SH12}. 
At a high in-plane wavevector $q$, the SS and AS$_2$ modes become  bound modes that are localized  around the interfaces; the field amplitudes take maxima at the interfaces and exponentially decay away from them. In the limit of $q\to\infty$, the energy dispersions for both of the SS and AS$_2$ modes become flat and  saturate to the same, finite frequency below $\omega_p$. Meanwhile, at $q=0$ the AS$_2$ mode starts from a frequency lower than that of the SS mode, and thus the former exhibits a larger dispersion. 
The SS mode can have a real longitudinal wavenumber $\kappa_\eta$ at low $q$ and thus it can be radiative, i.e., it can be coupled to light modes outside the cavity  when metal mirrors have finite thicknesses. As a result, the SS mode can contribute to optical properties of the system. With increasing $q$, the longitudinal wavenumber $\kappa_\eta$ changes from a real value to a purely imaginary value and the SS mode changes into the nonradiative mode. In contrast, the AS$_2$ mode is nonradiative over the entire in-plane momentum and can potentially be applied to nanoscale plasmonic waveguides owing to its localized and dispersive nature. The difference between two surface modes at low $q$ also results in their qualitatively different low-energy polarization behavior; the dipole moment of the SS mode points to the in-plane direction while that of the AS$_2$ mode is purely out-of-plane. 
 It is worthwhile to mention that in the context of plasmonics the nanoscale localization of such surface modes has previously been utilized for squeezing light waves below the diffraction limit \cite{NL06,MC14}.

The highest band (denoted as TM$_{2}$ in Fig.~\ref{fig6}(a)) starting from the plasma frequency $\omega_p$  at $q=0$ is essentially the upper branch of the standard plasmon polaritons. This hybridized mode reduces to the bare plasmon (photon) excitation in the limit of $q\to0$ ($q\to\infty$). The next highest band (denoted as TM$_{1}$ in Fig.~\ref{fig6}(b)) initiates from a frequency slightly below the plasma frequency and has a physical origin similar to the TM$_2$ mode, but there still exists a difference that the TM$_{1}$ mode sustains the nonvanishing  hybridization with the electromagnetic fields at low $q$. The mode functions ${\boldsymbol u}^\perp_\eta$ of the TM$_{1,2}$ modes and the corresponding dipole moments predominantly point to the directions parallel to the interfaces  at low $q$. The longitudinal wavenumber $\kappa_\eta$ remains real over the entire in-plane momentum space, and thus the amplitudes of the TM$_{1,2}$ modes extend over the insulator region (see also Fig.~\ref{figapp1} in Appendix~\ref{app2}).

When the plasma frequency is close to the phonon frequency and the hybridization with plasmons becomes particularly prominent, the above crossover behavior of the RTM$_0$ mode also manifests itself as the emergence of roton-type excitations. To see this explicitly, in Fig.~\ref{fig8} we show the energy dispersion of the RTM$_0$ mode at different light-matter coupling strengths $g$ with a low plasma frequency. It is evident that the energy dispersion takes the minimum value at finite in-plane momentum $q^{*}>0$, indicating the formation of roton excitations. Physically, this minimum roughly corresponds to the vanishing point of the longitudinal wavenumber  $\kappa_\eta=0$, at which the crossover from low to high in-plane momentum behavior occurs (compare Figs.~\ref{fig7}(c) with (d)). While a specific value of  the finite in-plane momentum $q^*$ depends on system parameters such as the thickness of the insulator, it is a general feature that roton excitations eventually disappear (i.e., $q^*\to 0$) as the plasmon component in the eigenmode becomes less significant by, for instance, increasing either the light-matter coupling strength $g$ or the plasma frequency $\omega_p$. 

\begin{figure}[t]
\includegraphics[width=50mm]{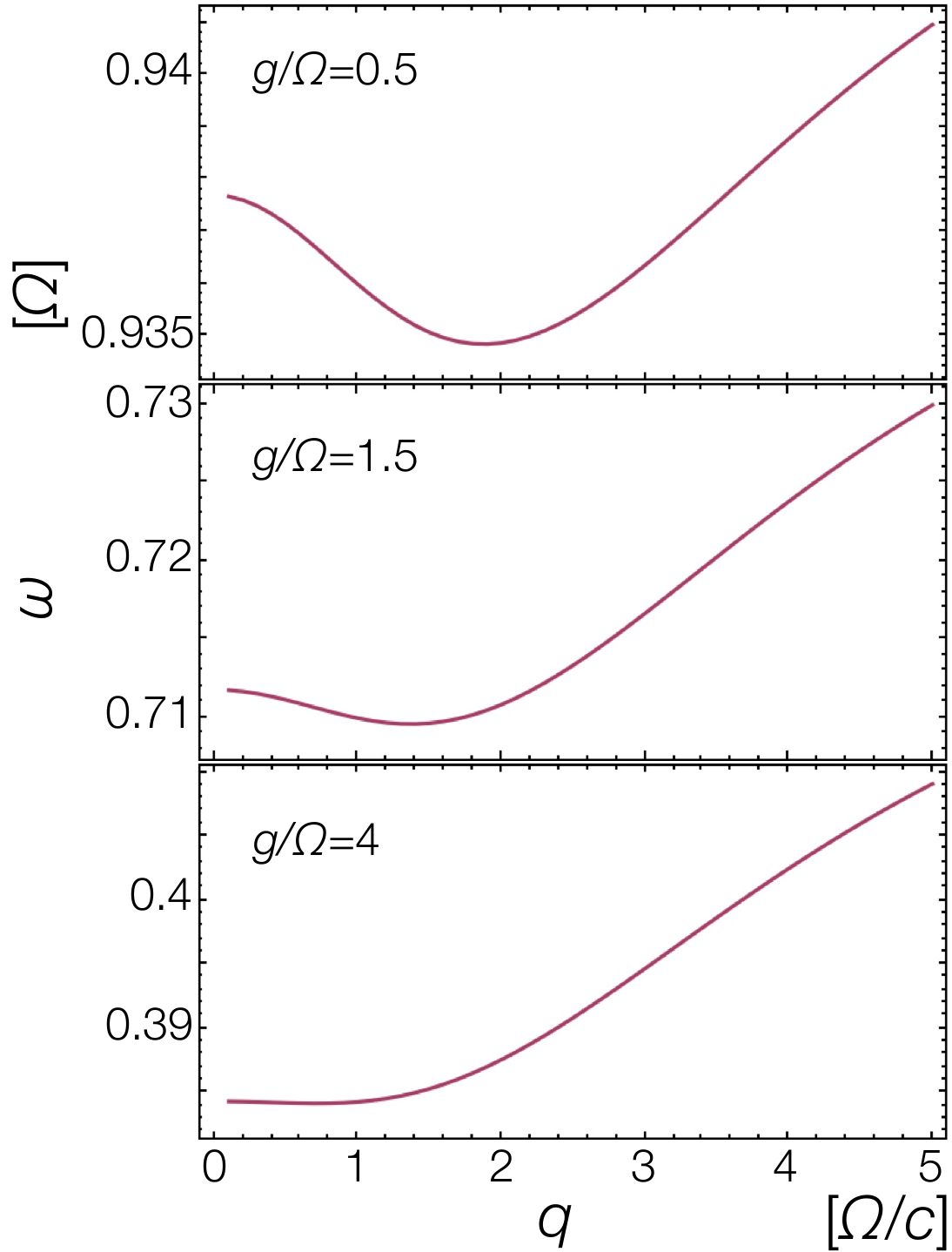} 
\caption{\label{fig8}
The RTM$_0$ mode at different light-matter coupling strengths $g$ with a low plasma frequency. The strong hybridization with plasmons leads to the emergence of the minimum of the energy dispersion at finite in-plane momentum $q^*>0$, indicating the formation of roton-type excitations. The parameters are $d=c/\Omega$ and  $\omega_p=2\,\Omega$. 
}
\end{figure}

The lowest mode in the TM polarization, which is denoted as AS$_1$, lies slightly below the lower bulk dispersion  (black dotted curve) and converges to the linear photon dispersion at $q\to 0$ in contrast to the other modes featuring quadratic asymptotes. This mode essentially shares the similar properties with the AS$_2$ mode; both of them are localized modes over the entire in-plane momentum $q$ and point out of the plane at low $q$ due to the surface charges.  The AS$_1$ mode asymptotically becomes a purely photon mode at $q\to 0$. 

The longitudinal modes L$_\pm$ in Fig.~\ref{fig6}(a) have the parity $P=\pm$ and lie at the constant frequency $\Omega^\parallel=\sqrt{\Omega^2+g^2}$. These nondispersive modes arise from the vanishing insulator permittivity and is characterized by the nonzero divergence of the mode function $\nabla\cdot{(e^{i\boldsymbol{q}\cdot\boldsymbol{\rho}}\boldsymbol f}^\parallel_\eta)\neq 0$. Both the in-plane and out-of-plane amplitudes of the longitudinal electric field extend over the insulator region  while the amplitudes vanish in the metal regions due to the boundary conditions. At low $q$, the out-of-plane component becomes dominant in the insulator region due to the surface charges.

Finally, we discuss the TE-polarized bands of both parities $P=\pm$ plotted in  Fig.~\ref{fig6}(c). The corresponding magnified plot around the phonon frequency is also shown in Fig.~\ref{fig6}(f). These TE modes have simpler structures than the TM ones, i.e., each of all the modes lies slightly above the higher or lower branches of the bulk dispersions (black dotted curves) due to energy enhancements by the cavity confinement. We label the higher set of the TE modes as TE$_{0,1}$ while the lower set of a large number of resonant TE modes as RTE$_{n=0,1,2\ldots}$. Because the mode functions (and thus the electric fields) for all the modes are continuous even at the interfaces, the origin of the enhanced energies in these TE modes can simply be understood as the acquisition of nonzero  longitudinal momentum associated with the cavity confinement. Indeed, the corresponding longitudinal wavenumber $\kappa_\eta$ remains real for all the TE modes over the entire region of  the in-plane momentum $\boldsymbol q$. This real-valuedness of $\kappa_\eta$ also indicates that the TE modes are essentially radiative photonic modes that can directly couple  to light modes outside the system when the thicknesses of the metal mirrors are taken to be finite. 

\section{Applications of cavity-based control of polariton dispersions  \label{app4}}
Substantial modifications of the polariton dispersions in the cavity geometry (see e.g., Fig.~\ref{fig6} in Sec.~\ref{sec3}) are generic features for collective dipolar modes coupled to confined light modes.  In this respect, the cavity-based control of energy-level structures can potentially be applied to a wide range of systems such as ionic crystals, molecular solids, assembled molecules, and excitonic systems. Organic molecules are particularly promising for such applications since their excitations have large dipole moments and crystals can achieve high-molecular densities \cite{LD98}. Recent experiments have demonstrated the strong coupling between a single molecule and a cavity mode even for the modest cavity quality factors \cite{CR16,BH17}. This means that the strength of the coupling between the molecular optical excitation and the cavity mode exceeds the decoherence rate of the excitations \cite{KY83,HJK98,HB16,FJ172}.   Besides a case study of the cavity-enhanced ferroelectricity in Sec.~\ref{sec4}, we discuss below that the cavity-induced changes in the hybridized dispersion can be utilized to improve the efficiency of organic light-emitting devices (LEDs) and to realize exciton-photon converters.

\begin{figure}[t]
\includegraphics[width=86mm]{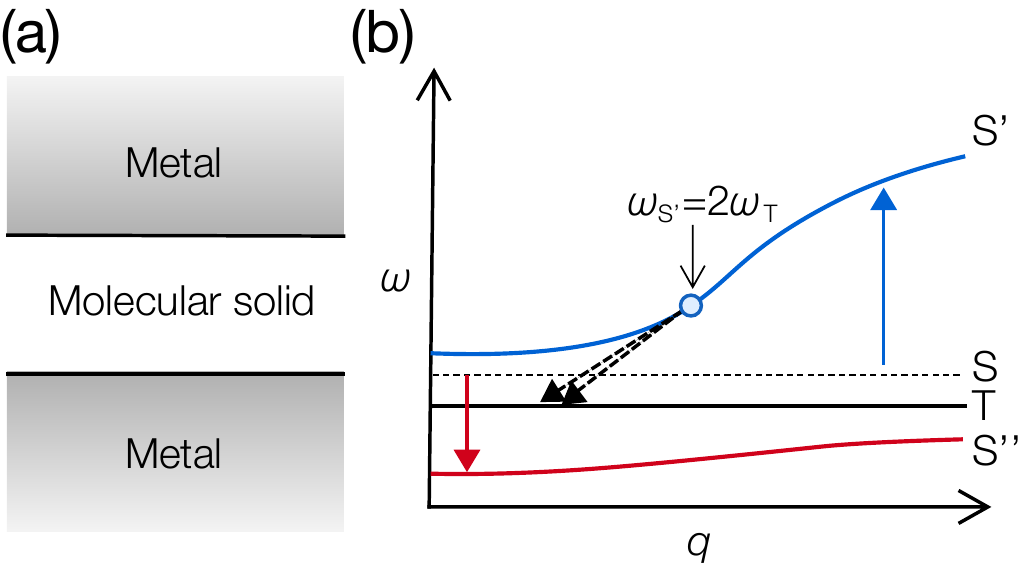} 
\caption{\label{figapp3}
Possible applications of the cavity setting to molecular and excitonic systems. 
(a) Schematic figure illustrating the proposed setup, in which molecular solids or assembled molecules interact with light modes confined between metal mirrors. (b) Schematic figure illustrating applications of the elementary excitations in the cavity setting to an organic light-emitting device and a photon-exciton conversion. The black horizontal solid (dotted) line denoted as T (S) indicates a bare nondispersive triplet (singlet) energy of a molecular state or an electron-hole pair. Only the singlet state S couples to confined light modes, resulting in the split into the multiple branches of the in-plane dispersions. A lower dispersion (red solid curve denoted as S$''$) can potentially be applied to enhance the efficiency of organic light-emitting devices. Tuning the in-plane momentum $q$ or the thickness of the layer, a higher dispersion (blue solid curve denoted as S$'$) can be used to realize the resonance condition $\omega_{{\rm S}'}=2\omega_{\rm T}$, leading to a conversion of a photon into a pair of triplet excitons as denoted by two dashed black arrows.   
}
\end{figure}

 We consider a crystal of polar molecules surrounded by metallic electrodes as shown in Fig.~\ref{figapp3}. The relevant infrared active mode can be either an electronic excitation, such as  the E1 electronic transition, or a vibrational excitation of the molecular crystal. While these two types of excitations have very different frequencies (i.e., optical vs mid-infrared), the general formalism presented in Sec.~\ref{sec3} should be applicable to both cases. Either one of these modes can be  introduced as a nondispersive mode having the frequency $\Omega$ in Eq.~\eqref{linmat}. To be concrete, we focus on the case of electronic excitations, and argue that the proposed cavity configuration can be applied to improve efficiencies in molecular LEDs.

A major challenge in organic LEDs is that the lowest excitonic state is a spin triplet, which is optically dark (i.e., it recombines through nonradiative channels). The reason for this is that Hund's rule for electrons in a molecule favors the spin alignment and thus the spin triplet excitonic states are lower in energy than the singlet ones.  Having a practical method of reversing the order of excitonic states by making the singlet state of an exciton lower in energy will have important implications for improving the efficiency of organic LEDs \cite{SK18}. 
In the cavity configuration, since triplet excitonic states are dark, they do not couple to light and are not strongly modified. In contrast, the singlet excitonic state hybridizes with confined light modes and forms the softened collective modes such as the RTM$_0$ mode shown in Fig.~\ref{fig6}(a). This should make it possible to have the lowest excitonic state as a spin singlet, thus realizing efficient organic LEDs (see the red solid curve in Fig.~\ref{figapp3}(b)). 
While this type of hybridizations have been discussed in the literature before \cite{SK18}, the novelty of the present consideration is  the inclusion of the continuum of both collective matter excitations and cavity light modes in a slab of a finite thickness as well as taking into account the dispersive nature of the hybridized excitations there. In general, the dispersive nature could be utilized to further tune the selective control of chemical reactivity landscapes in order to enhance or suppress certain types of photochemical reactions. Remarkably, an inversion of the singlet-triplet polaritons has recently been demonstrated in organic microcavities \cite{Eiznereaax4482}; it is interesting to consider whether such a technique can be transferred to continuum collective excitations associated with nontrivial dispersions as discussed here.

We next discuss a complementary idea of using a cavity to facilitate the conversion of photons into pairs of triplet excitons. This process could be useful for generating entangled pairs of excitations. At ambient conditions, such conversion process is challenging to realize because the resonant absorption of light occurs at the energy of singlet excitons and this energy is only slightly above the energy of the triplet excitons; typical energy spacing is characterized by the Hund's splitting and is an order of a few meV. 

This limitation can be overcome by utilizing the proposed cavity setup. As discussed before, when a molecular solid is confined in the cavity, the singlet exciton hybridizes with light modes while the triplet one does not. This can lead to the formation of a  hybridized singlet mode above the original exciton energy (see e.g., the blue solid curve S' above the black dashed  horizontal line S in Fig.~\ref{figapp3}(b)). One may, for example, use the SS(=TM$_0$) and  AS$_2$ modes for this purpose. Owing to the nontrivial dispersion of such a hybridized mode, one can tune the in-plane momentum $q$ (and the thickness $d$ or the plasma frequency $\omega_p$ if necessary) such that a hybridized singlet energy $\omega_{\rm S'}$ becomes equal to twice of the triplet energy $2\omega_{\rm T}$ and thus the conversion process of a single photon into two triplet excitons becomes resonant (cf. the dashed black arrows in Fig.~\ref{figapp3}(b)). One can create singlet exciton-polaritons at finite in-plane momenta by, e.g., having incident light at  nonzero angle or  constructing a grating on the surface of the metals.

\section{Details of the variational analysis  \label{app_vari}}

\begin{figure}[b]
\includegraphics[width=60mm]{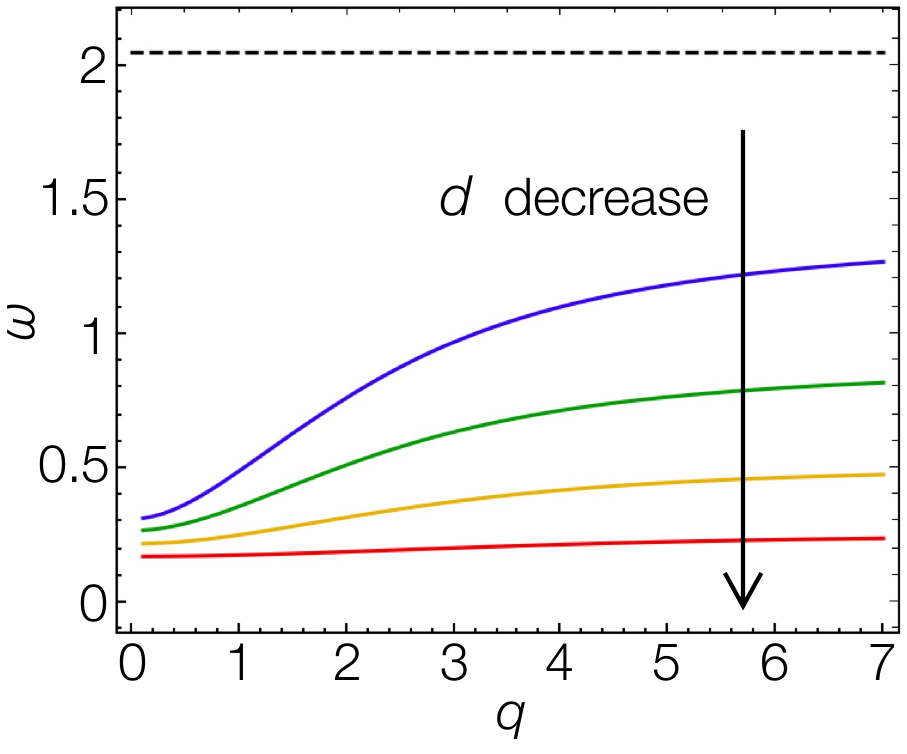} 
\caption{\label{fig10}
Softening of the phonon mode relevant to the paraelectric-to-ferroelectric phase transition, which is labeled as the RTM$_0$ mode in Fig.~\ref{fig6}. The solid colored curves represent the in-plane dispersions of the softened phonon mode with decreasing the insulator thickness $d$ from top to bottom curves. Each dispersion is obtained by setting the phonon-frequency parameter $\Omega^2$ equal to the effective value $r_{\rm eff}^*$ determined from the variational analysis for each thickness (cf. Fig.~\ref{fig9}(a)). The black dashed line represents the effective phonon frequency in the bulk case.
We set $c=\hbar=\beta=1$, $\omega_p=5$, $U=1$, $g=3.5$, and $\Lambda_c=10^2$. The  tuning parameter is fixed to be $r=-45$. From top to the bottom solid curves, the thickness is varied from $d=4$ to $d=1$ with step $\delta d=1$.
}
\end{figure}

We here provide technical details about the variational analysis. As explained in the main text, we minimize the following variational free energy with respect to a variational parameter $r_{\rm eff}$ (cf. Eq.~\eqref{varmain}):
\eqn{F_{v}=-\frac{1}{\beta}\ln Z_{0}+\delta E_{{\rm eff},1}+\delta E_{{\rm eff},2}.}
Here, the first term corresponds to the free energy for the noninteracting theory with 
\eqn{\label{appvar1}
Z_{0}\left(r_{{\rm eff}}\right)=\prod_{\boldsymbol{q}n\lambda}\left[1-e^{-\beta\omega_{\boldsymbol{q}n\lambda}\left(r_{{\rm eff}}\right)}\right].
}
The second term $\delta E_{{\rm eff},1}$ contributes from an expectation value of the quadratic terms in the energy difference $\hat{H}_{\rm tot}(r)-\hat{H}_{{\rm tot},0}(r_{\rm eff})$ with respect to the variational state, resulting in
\eqn{\label{appvar2}
\delta E_{{\rm eff},1}\left(r,r_{{\rm eff}}\right)&=&\nonumber\\
\frac{\hbar\left(r-r_{{\rm eff}}\right)}{4\sqrt{r_{{\rm eff}}}}\sum_{\boldsymbol{q}n\lambda}&&\!\!\!c_{\boldsymbol{q}n\lambda}\coth\left(\frac{\beta\hbar\omega_{\boldsymbol{q}n\lambda}(r_{{\rm eff}})}{2}\right).
}
Here, positive coefficients $c_{\boldsymbol{q}n\lambda}$ are defined by
\eqn{c_{\boldsymbol{q}n\lambda}\equiv\zeta(\omega_{\boldsymbol{q}n\lambda})\int_{-d/2}^{d/2}dz\,\left|\boldsymbol{f}_{\boldsymbol{q}n\lambda}(z)\right|^{2},
}
where $\zeta(\omega)=g^4/[g^4+(r_{\rm eff}-\omega^2)^2]$.
The mode function $\boldsymbol{f}_{\boldsymbol{q}n\lambda}(z)$ represents $\boldsymbol{f}_{\boldsymbol{q}n\lambda}^\perp$ for transverse modes with $\Lambda={\rm TM},{\rm TE}$, while it represents $\boldsymbol{f}_{\boldsymbol{q}n\lambda}^\parallel$ for longitudinal modes with $\Lambda={\rm L}$; their explicit functional forms are given in Appendix~\ref{app2}.
The final term $\delta E_{{\rm eff},2}$ results from the nonlinear term in the  microscopic Hamiltonian; after using the Cauchy-Schwartz inequality, we can bound it as
\eqn{\label{appvar3}
\delta E_{{\rm eff},2}\left(r_{{\rm eff}}\right)\leq\nonumber\\
\frac{\hbar^{2}U}{4M^{2}r_{{\rm eff}}N}\!&&\left[\!\sum_{\boldsymbol{q}n\lambda}\!\tilde{c}_{\boldsymbol{q}n\lambda}\coth\left(\frac{\beta\hbar\omega_{\boldsymbol{q}n\lambda}(r_{{\rm eff}})}{2}\right)\right]^{2}\!\!\!,
} 
where we introduce positive coefficients
\eqn{
\tilde{c}_{\boldsymbol{q}n\lambda}\equiv\zeta(\omega_{\boldsymbol{q}n\lambda})\sqrt{d\int_{-d/2}^{d/2}dz\,\left|\boldsymbol{f}_{\boldsymbol{q}n\lambda}(z)\right|^{4}}.
}
Summing up the right-hand sides of Eqs.~\eqref{appvar1}, \eqref{appvar2}, and \eqref{appvar3}, we obtain the variational free energy. While strictly speaking this gives the upper bound of Eq.~\eqref{varmain}, in this paper we term this summation as the variational free energy $F_v$ for the sake of notational simplicity.

One possible way to infer the quality of the present variational theory is to check the convexity of the variational free energy and its variance around the global minimum $r_{\rm eff}=r_{\rm eff}^*$. In general, when the free-energy landscape turns out to be nonconvex and has several local optima, it should raise the question about the validity of a choice of variational states. However, in the present analysis, the free-energy landscape is very well behaved and has only a single global minimum. In particular, the landscape around the global minimum becomes sharper as we approach to the transition point at $r_{\rm eff}^*=0$. These features can be inferred from Fig.~\ref{fig9}(a) in the main text.  

In addition to plotting the variational free energies and the resulting phase diagrams as shown in Fig.~\ref{fig9}, it is also useful to plot the most-softened phonon mode  (i.e., the RTM$_0$ mode) with varying the insulator thickness $d$, but at the fixed tuning parameter $r$ (see Fig.~\ref{fig10}). It is this softening that destabilizes the paraelectric phase when it goes down to zero energy, triggering the phase transition. Thus, the unstable dipolar phonon mode at the verge of the transition is precisely the transverse one and points to the in-plane direction (cf. Fig.~\ref{fig7}(a,c)).
We emphasize that the results in  Fig.~\ref{fig10} are obtained by using the renormalized effective phonon frequency $\Omega_{\rm eff}^*=\sqrt{r_{\rm eff}^*}$ for each thickness $d$, which is determined from identifying the minimum of the variational free energy  including the nonlinear interaction in a self-consistent manner (cf. Fig.~\ref{fig9}(a)). We note that all the contributions from the `spectator' modes, i.e., the modes other than the RTM$_0$, are also included in the variational analysis to determine the renormalized phonon frequency. As shown in Fig.~\ref{fig10},  to realize the ultimate softening of the RTM$_0$ mode, a sparser band structure as realized with a thinner layer is preferable since the phonon hardening due to the nonlinear repulsive interaction between the energetically dense elementary modes can be mitigated. 

\bibliography{reference}

\end{document}